\newcommand{\Gaia}{\emph{Gaia}}
\newcommand{\GDR}{\Gaia\ DR}
\newcommand{\G}{$G$}
\newcommand{\BP}{$G_{\rm BP}$}
\newcommand{\RP}{$G_{\rm RP}$}
\documentclass{aa}  

\usepackage{subfig}
\usepackage{graphicx}
\usepackage{color}
\usepackage{txfonts}
\usepackage[normalem]{ulem}
\usepackage{cancel}
\usepackage{float}
\usepackage{hyperref}
\begin{document} 

   \title{Gaia Data Release 1}

   \subtitle{Validation of the photometry}

   \author{
  D.~W.~Evans\inst{1}
\and
  M.~Riello\inst{1}
\and
  F.~De~Angeli\inst{1}
\and
  G.~Busso\inst{1}
\and
  F.~van~Leeuwen\inst{1}
\and
  C.~Jordi\inst{2}
\and
  C.~Fabricius\inst{2}
\and
  A.~G.~A.~Brown\inst{4}
\and
  J.~M.~Carrasco\inst{2}
\and  
  H.~Voss\inst{2}
\and
  M.~Weiler\inst{2}
\and
  P.~Montegriffo\inst{3}
\and
  C.~Cacciari\inst{3}
\and
  P.~Burgess\inst{1}
\and
  P.~Osborne\inst{1}
}

\institute{
Institute of Astronomy, University of Cambridge, Madingley Road, Cambridge
CB3 0HA, UK\\
\email{dwe@ast.cam.ac.uk}
\and
Institut de Ci\`encies del Cosmos, Universitat de Barcelona (IEEC-UB),  Mart\'i Franqu\`es 1, E-08028 Barcelona, Spain
\and
INAF -- Osservatorio Astronomico di Bologna, via Ranzani 1, 40127 Bologna, Italy
\and
Sterrewacht Leiden, Leiden University, PO Box 9513, 2300 RA Leiden, the Netherlands
}

\date{Received \textbf{Month Day, 201X}; accepted \textbf{Month Day, 201X}}

 
\abstract
{}
{The photometric validation of the \GDR1 release of the ESA \Gaia\ mission is described and the quality of the data shown.}
{This is carried out via an internal analysis of the photometry using the most constant sources. Comparisons with external photometric catalogues are also made, but are
limited by the accuracies and systematics present in these catalogues. An analysis of the quoted errors is also described. Investigations of the calibration coefficients reveal some 
of the systematic effects that affect the fluxes. } 
{The analysis of the constant sources shows that the early-stage photometric calibrations can reach an accuracy as low as 3 mmag.} 
{}


\maketitle
%


\section{Introduction}

The photometric calibration of the first data release of the \Gaia\ catalogue \citep{GDR1} aims to  achieve mmag-level
precision \citep{PhotTopLevel}. This is carried out via an internal, self-calibrating method  as detailed in \citet{PhotPrinciples}. No comparison with a set of standards would be sufficient to confirm that the accuracies quoted for the photometry
were valid{} because the precision aimed for is better than the precision of other  currently available large catalogues of 
photometric standards. Also, unexpected systematic effects have been found in the \Gaia\ data that required additional 
calibrations to be carried out with respect to the initial plan \citep{PhotProcessing}, such as linear trends with 
time 
and increased background level.
It is necessary to 
check 
that these calibrations have removed all the
systematic effects and that the accuracies achieved are close to the photon noise level. It is expected that future releases of the
\Gaia\ catalogue will have improved accuracies as further calibrations are introduced into the data processing.

Although no colour nor spectral information is included in this data release, some validation is given to the processing
of the spectral calibrations of  blue and red
photometers (BP and RP, respectively). This is due to the use of colour information in the calibration
of the G-band photometry, which is itself internally calibrated.  We also present some results of the  validation  of the epoch photometry
available in the \Gaia\  EPSL release \citep{VariEpsl}.

Sections \ref{Sect:BpRpCal} to \ref{Sect:Convergence} of this paper cover the direct validations of the calibrations carried out in the  photometric processing.
 This is followed by internal consistency checks using the accumulated photometric data, 
the photometric residuals, and an analysis of the epoch photometry of constant sources in Sects. \ref{Sect:Accumulation} -  \ref{Sect:Constant}. 
External comparisons are then described  in  Sect. \ref{Sect:External}. Finally, Sect.~\ref{Sect:Conclusions} summarises  the conclusions. 
 Appendix~\ref{acronyms} contains a list of the acronyms used in this paper.

Further validation of the overall catalogue can be found in \citet{CU9Validation}.

The following subsection provides a brief description of the \Gaia\ instruments and data. Many more details are available 
from \citet{GaiaMission}.

\section{Input data}

 \Gaia\ is a scanning satellite. The full sky is expected to be covered in about six months of observations, but 
the number of observations per source largely depends on the astrophysical coordinates of the source.

The main input data for the photometric processing comes from the Astrometric Field (AF) CCDs.
This is an array of seven rows (parallel to the along-scan (AL) direction) and nine strips (parallel to the across-scan (AC) direction) 
of CCDs collecting light in the \Gaia\ \G\ broad band. 
Colour information for each source (also a fundamental ingredient for the photometric calibrations) 
is derived from the low-resolution spectra collected by the BP and RP instruments.
The light is dispersed in the along-scan direction.
 
In the following we will refer to field-of-view (FoV) and CCD transits: a FoV transit includes
several CCD transits (usually nine AF, one BP, and one RP CCD transit).
It should be noted that the two FoVs are simultaneously projected onto the focal plane array. 

Only small windows centred on the detected sources are downloaded from the satellite. The size of these 
windows depends on the magnitude estimated on board; only bright sources are observed with 2D windows. 
Different configurations are referred to as ``window classes''.
The shape of the windows (normally rectangular) can be complicated by conflicts between adjacent sources 
in crowded regions in the sky. These non-nominal cases have not been treated yet and have  not contributed to the 
photometry published in \GDR1.

The CCDs are operated in a time-delayed integration (TDI) mode whereby charges are integrated while they 
move across the CCD. The effective
exposure time over one CCD is approximately 4.5 seconds, but this can be reduced by activating the 
``gates''. Several different gate configurations are defined and a particular gate is assigned to each 
CCD transit depending on the magnitude of the source as estimated from the strip of CCD preceding 
the AF (known as the  ``star mapper'' or SM CCDs) and the AC position of the source in the focal plane.
Over a FoV transit, different gate configurations can be used on individual CCD observations. The activation of a gate 
triggered by the transit of a bright source, will affect all other sources observed simultaneously
in the same region of a CCD. It may also affect only part of a window, thus creating complex gate cases 
that have not yet been treated  by the photometric processing.

Different gate and window class configurations  effectively define different instruments (referred to as
calibration units) that need to be calibrated to form one consistent reference system \citep[for more details see][]
{PhotPrinciples}.

The input data to the photometric processing consists of image parameters (such as fluxes, centroids, and 
goodness of fit measurements) for the SM and AF CCD transits, and raw BP and RP spectral data. Errors on the
G-band flux measurements are estimated in the image parameter determination (IPD) process; for more details 
refer to \citet{IdtRef}.

Two  significant and unexpected features were discovered during the commissioning period and required
the introduction of ad hoc calibrations. One is the presence of stray light scattered by the solar shield,
causing the background level to be up to 
two orders of magnitude higher than
expected (with large variations depending on the rotation phase of the satellite). The additional
stray light component of the background can be calibrated, but the associated noise will affect 
performance for faint objects. The other feature is a decrease over time of the throughput of the instruments
due to continued contamination by water ice.  
The wavelength-dependent transmission loss is different for the two FoVs and varies across
the focal plane. This adds a systematic effect to the photometric data that is orders of magnitude larger
than expected and in particular affects our ability to create a consistent reference catalogue for the
internal calibration. An additional calibration of this strong time dependency in the transmission 
had to be included to solve this problem. 

\section{Validation of BP/RP spectral calibrations}\label{Sect:BpRpCal}

\newcommand{\fda}[1]{{\bf #1}}

Even though colour information is not included in Gaia~DR1, BP and RP data are processed to produce the colour 
information required to calibrate the G-band photometry. This section focusses  on the validation of the along-scan geometric calibration of the spectrophotometric data. This is a fundamental element in the computation 
of colour information in the form of spectrum shape coefficients \citep[SSC; see][]{PhotPrinciples}. The application of this calibration and of the 
nominal dispersion function brings all the data onto the same wavelength system.

The BP/RP windows that are  assigned on board will generally not be well centred on the source. This is expected, given 
the design of the instrument, and  is due to various 
factors: the location of the centroid from the SM observation may not be very accurate, and sources may have a 
non-negligible motion along or across scan. This implies that considering 
an arbitrary reference wavelength, the location  will not correspond to the same location in sample space even in 
spectra of the same source \citep[for a definition of sample space see][]{GaiaMission}.

The location of the centre of the source, or more precisely of a reference wavelength in the dispersed image
of the source, can be predicted by extrapolation from the series of source centroids of each of the AF observations
that precede the BP/RP observations  in a FoV transit. However, this requires an accurate knowledge of the
geometric calibration of the BP/RP CCDs with respect to the AF CCDs (and of the satellite attitude).

To calibrate the geometry of the instrument, this prediction needs to be compared to the actual location 
of the reference wavelength in the observed spectra. This is done by selecting a small fraction of the observed
spectra based on a filter in colour and magnitude, so that we can be confident that we are using spectra 
of sources with a similar spectral energy distribution, where the sample position corresponding to the reference 
wavelength will be the same (except for the effect of non-perfect centring of the window mentioned above) and
that we are filtering observations with a high S/N.
The colour range adopted is $[0.3, 0.6]$ for the calibration of the BP instrument and $[1.3, 1.6]$ for the calibration of the
RP one. These correspond approximately to spectral types F and K (based on
nominal knowledge of the instrument and pre-launch simulations). This colour
selection, in addition to other filters designed to select isolated spectra
and to avoid spectra that are affected by cosmic rays, yields a sufficient
number of calibration spectra. This is of the order of several hundreds for each calibration unit of the  large-scale component of the AL  geometric
calibration model which is the one that is updated most often (every 20 OBMT revolutions or about 5 days for \GDR1). 
The selected spectra are aligned and used to generate a reference spectrum, which is then fitted back to each spectrum 
to evaluate the sample position of the reference wavelength within the actual sampling. 
 Two reference spectra are defined for the entire dataset, one for BP and one for RP.

At this stage, the processing has concentrated on differential calibrations of the various instrument
configurations onto the same internal reference system. This is then tied to the absolute system adopting the
nominal pre-launch knowledge of the instrument. In this simplified schema, it is acceptable to adopt as the reference
wavelength the nominal value which corresponds to the central sample of a perfectly centred window and to assume that 
the reference spectrum (being the result of an accumulation over an extremely large number of observed spectra) will be
representative of a perfectly centred spectrum.

The geometric calibration model is defined by the following components \citep[for more details see][]{PhotPrinciples}:

\begin{itemize}
\item a large-scale component, computed over a short timescale, defined by a linear combination of shifted Legendre
polynomials describing overall effects of translation, rotation, and curvature;
\item an offset for different gate configurations (relative to the un-gated case) computed on a longer timescale, 
taking into consideration the residual effects;
\item an offset for different CCD AC stitch blocks, also computed on a long timescale, taking into account the effects
due to the photolithography process used to manufacture the CCDs \citep[for more details on the definition
of the stitch blocks see][]{GaiaMission}.
\end{itemize}

\begin{figure} \resizebox{\hsize}{!}{\includegraphics[angle=270]{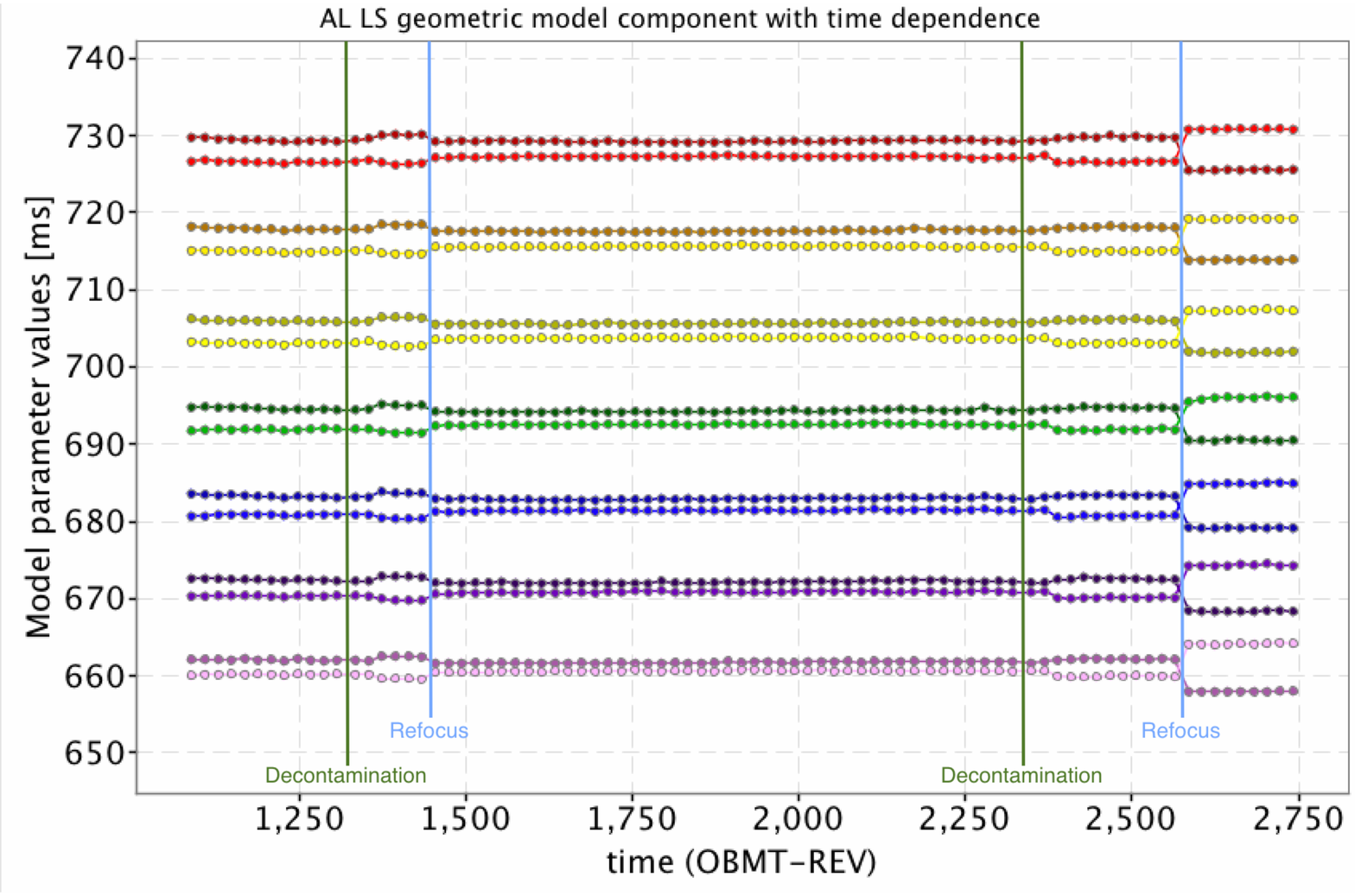}} \caption{Evolution in time
of the zeroth-order coefficients of the large-scale component of the BP geometric calibration. The units on the ordinate
axis are  ms (the TDI period is 1 ms). 
 The units on the abscissa axis are OBMT revolutions (one revolution corresponds to approximately  6 hours). The 
OBMT range covers the entire science acquisition period for \GDR1, i.e. between 25 July 2014 and 16 September 2015. Different 
colours are used to indicate different 
CCD rows (CCD rows 1 to 7 from red to violet, lighter colours for the preceding field of view, darker ones for the following field of view). Each large-scale calibration
covers a time range of about 20 revolutions (5 days).}
\label{Fig:bpAlGeoCalLs0Zoom}
\end{figure}

Sudden variations in the values of the calibration coefficients over time should only take place
corresponding to particular and known satellite events or features/changes in the input data produced by the
upstream systems.
Therefore, the main validation analysis is based on the temporal evolution of the calibrations.
Figure \ref{Fig:bpAlGeoCalLs0Zoom} shows the evolution versus time \citep[in OBMT revolutions, where one revolution lasts 
approximately 6 hours; see][]{GaiaMission} of the zeroth order coefficients
in the large-scale component. Some known events are marked in the plot using vertical lines (two decontamination 
activities in dark green and two refocus activities in blue). As expected, these significantly affect the calibrations. 
Decontamination activities were introduced to mitigate the problem of contamination affecting mirrors and CCDs. During these
activities, the mirrors and CCDs were heated. The decontamination and refocus  activities mainly affect  the basic
angle, while they seem to have an almost negligible effect on the relative geometry of AF and BP/RP.

It is interesting to note that the variations following a decontamination
seem to take place with some delay. This is likely  due to the fact that data collected just after a decontamination
event are not of sufficient quality
to generate an updated AF geometric calibration, and therefore this update takes 
place only once the entire instrument has cooled down sufficiently. Variations in the level of the BP/RP geometric calibration 
coefficients are expected whenever a new geometric calibration for the AF field comes in place. This occurs because the extrapolation of the AF
centroids to the BP/RP CCDs depends on the AF geometric calibration.

As can be seen from Fig. \ref{Fig:bpAlGeoCalLs0Zoom}, the large-scale component is very stable over stretches of
nominal operations. 

\begin{figure} \resizebox{\hsize}{!}{\includegraphics[angle=270]{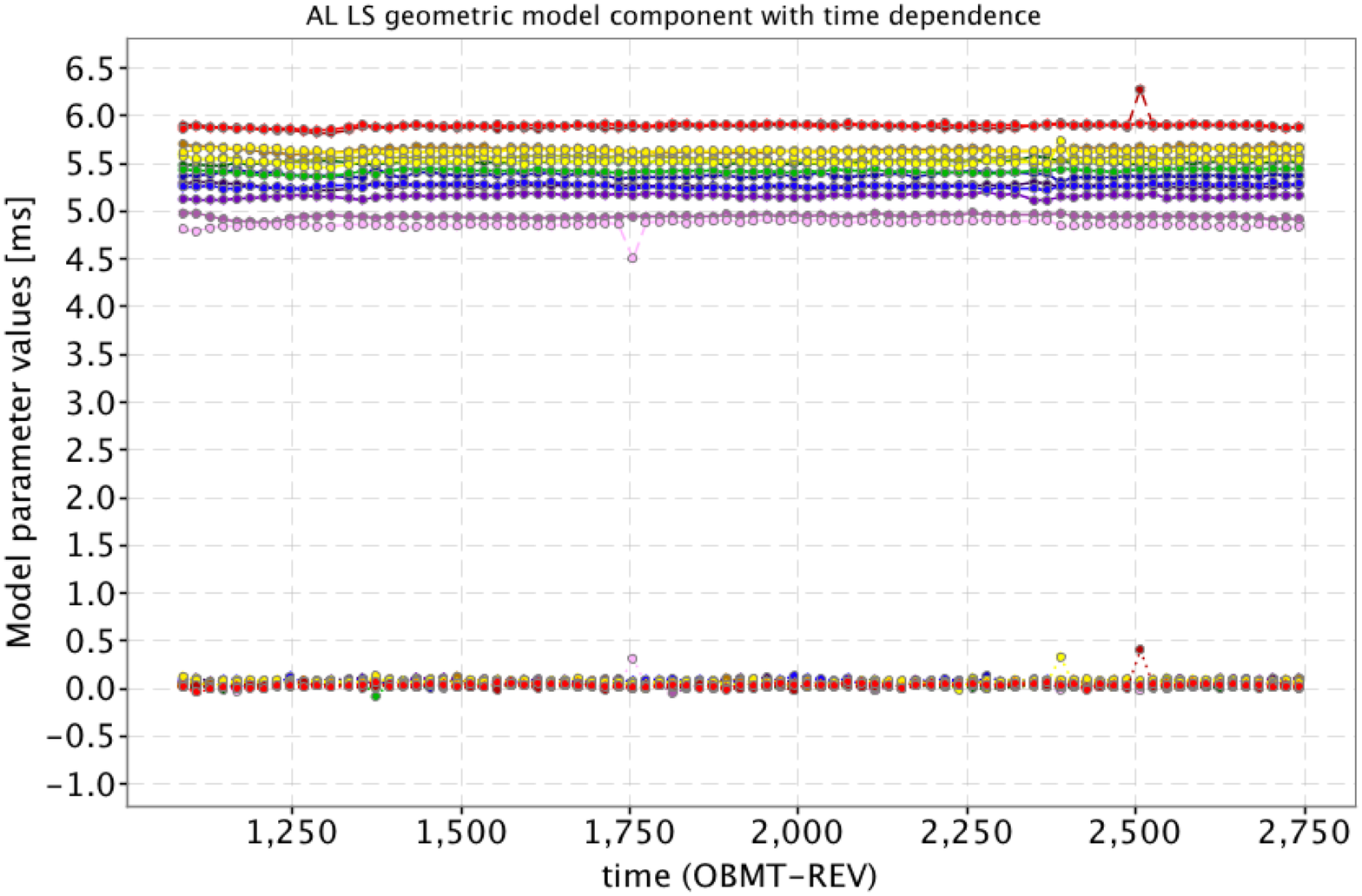}} \caption{Evolution in time
of the first- (distributed between 4.5 and 6 ms) and second-order coefficients (distributed between $-$0.5 and 0.5 ms)  of the 
large-scale component of the BP geometric calibration.  The units on the ordinate
axis are  ms. 
{ The units on the abscissa axis are OBMT revolutions (one revolution corresponds to about 6 hours).}
{The OBMT range covers the entire science acquisition period for \GDR1.}
Colour-coding is as in Fig. \ref{Fig:bpAlGeoCalLs0Zoom}. }
\label{Fig:bpAlGeoCalLs1And2Zoom}
\end{figure}

Figure \ref{Fig:bpAlGeoCalLs1And2Zoom} shows the time evolution of the first- and second-order large-scale coefficients.
The colour-coding is the same as in Fig. \ref{Fig:bpAlGeoCalLs0Zoom}. The second-order coefficients are always very 
close to 0. Both sets of coefficients are quite stable.

\begin{figure} 
\resizebox{\hsize}{!}
{\includegraphics[angle=270]{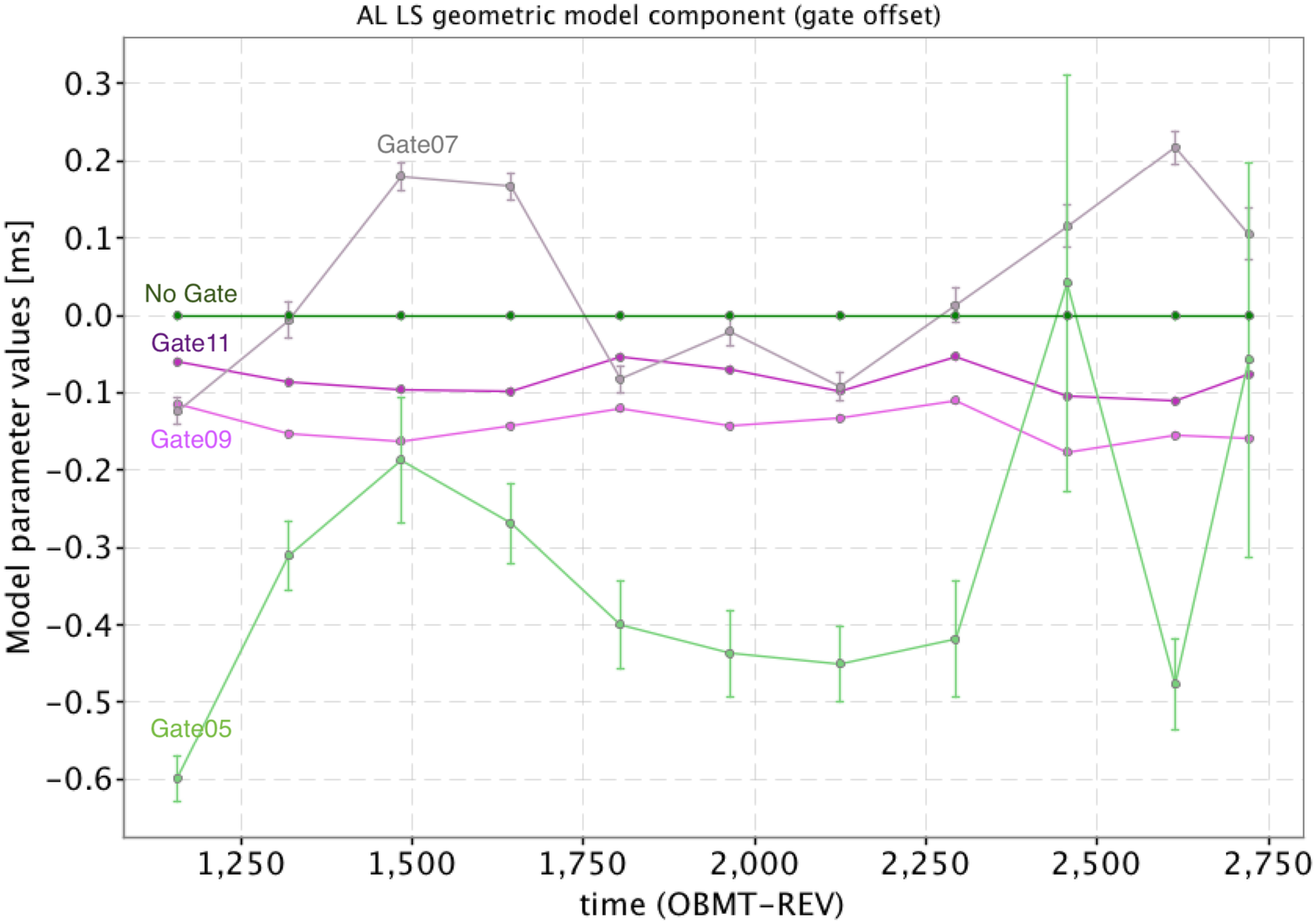}} 
\caption{Evolution in time
of the gate offset coefficients of the large-scale component of the BP geometric calibration. The units on the ordinate
axis are ms. 
 The units on the abscissa axis are OBMT revolutions (one revolution corresponds to about 6 hours).
The OBMT range covers the entire science acquisition period for \GDR1.
Different colours are used to indicate different gate configurations  as indicated by the labels. }
\label{Fig:bpAlGeoCalGateOffsetZoom}
\end{figure}

Finally, Fig. \ref{Fig:bpAlGeoCalGateOffsetZoom} shows the offset calibrated for different gate configurations.
 The only gates that could be calibrated over the whole period are Gate09,  Gate11, Gate07, and Gate05. The coefficients for Gate05 and Gate07 are 
quite noisy (the width of the distribution of the parameter values for these two configurations is 0.22 and 0.14 pixels to be compared with 0.02 obtained for both Gate09 and Gate11). 
This is due to the small amount of data available for these calibrations. 
Time ranges covering about 160 revolutions (40 days) were used for this run of the gate offset calibration.
Longer time ranges  could be adopted in future runs if the calibrations are sufficiently stable. 

The final set of coefficients, calibrating small-scale effects on the scale of CCD AC stitch block, produces
offsets that are always below 0.05 pixel in absolute value and are quite stable (the width of the distribution over time
of these parameters is lower than 0.02 pixels in every calibration unit).

The validation of the geometric calibrations  is primarily based on the analysis 
of the standard deviation of the single calibrations. 
The standard deviation of the zeroth-order parameter for each single calibration unit, over the 6 months 
following the first refocus event, is of the order of $0.06$ ms for BP (equivalent to $0.06$ pixel or $0.5$ nm in 
terms of wavelength) and $0.15$ ms for RP (i.e. $0.15$ pixel or $1.65$ nm in terms of wavelength).
Errors of this size are negligible when computing the spectrum shape coefficients used for the photometric 
calibrations. This is the only relevant quantity for \GDR1 as spectral data is not yet included in the release.

It should be noted that systematic errors on the geometric calibration parameters would not affect the photometric calibrations
as they will simply result in a slightly different set of SSC bands being used for the definition of the colour 
information.

The RP results (not shown in this paper) are equivalent to the BP ones.

\section{Validation of BP/RP stray light calibration}

As described in \citet{PhotProcessing}, the current implementation of the background correction takes into account only the stray light calibration as it is the most important contribution to the background. 
As  \citet{PhotPrinciples} have noted, this correction of the stray light also includes the smoother component of the astrophysical background.
The stray light is modelled as a discrete 2D map, obtained by accumulating eight revolutions of data (corresponding to roughly  2 days). The map coordinates are the heliotropic coordinate spin phase and the AC coordinate. The 1D and 2D transits are processed separately because analysis of the data shows that while the structure
 of the map is  very similar, there is a small offset (still under investigation)  between the two. 
Since there are many more 1D transits than 2D transits, 
they have a much higher weight in the determination of the bin values, and using a map built with both kinds of observations leads to an overcorrection when removing the background for the 2D transits. 
 The amplitude of this offset depends on the CCD and varies between 0 and 2 electron/pixel/s. This does not affect the photometry because it is calibrated out by creating separate maps for 1D and 2D
windows.  
In addition,  the grid used to build the maps for 1D and 2D transits are different, with 360 bins in phase and 20 in {AC coordinate} in the former  and 180 and 15 in the latter. This  allows fewer empty bins for the maps built with 2D transits.

\begin{figure*}
\centering
\includegraphics[width=0.4\textwidth,angle=0]{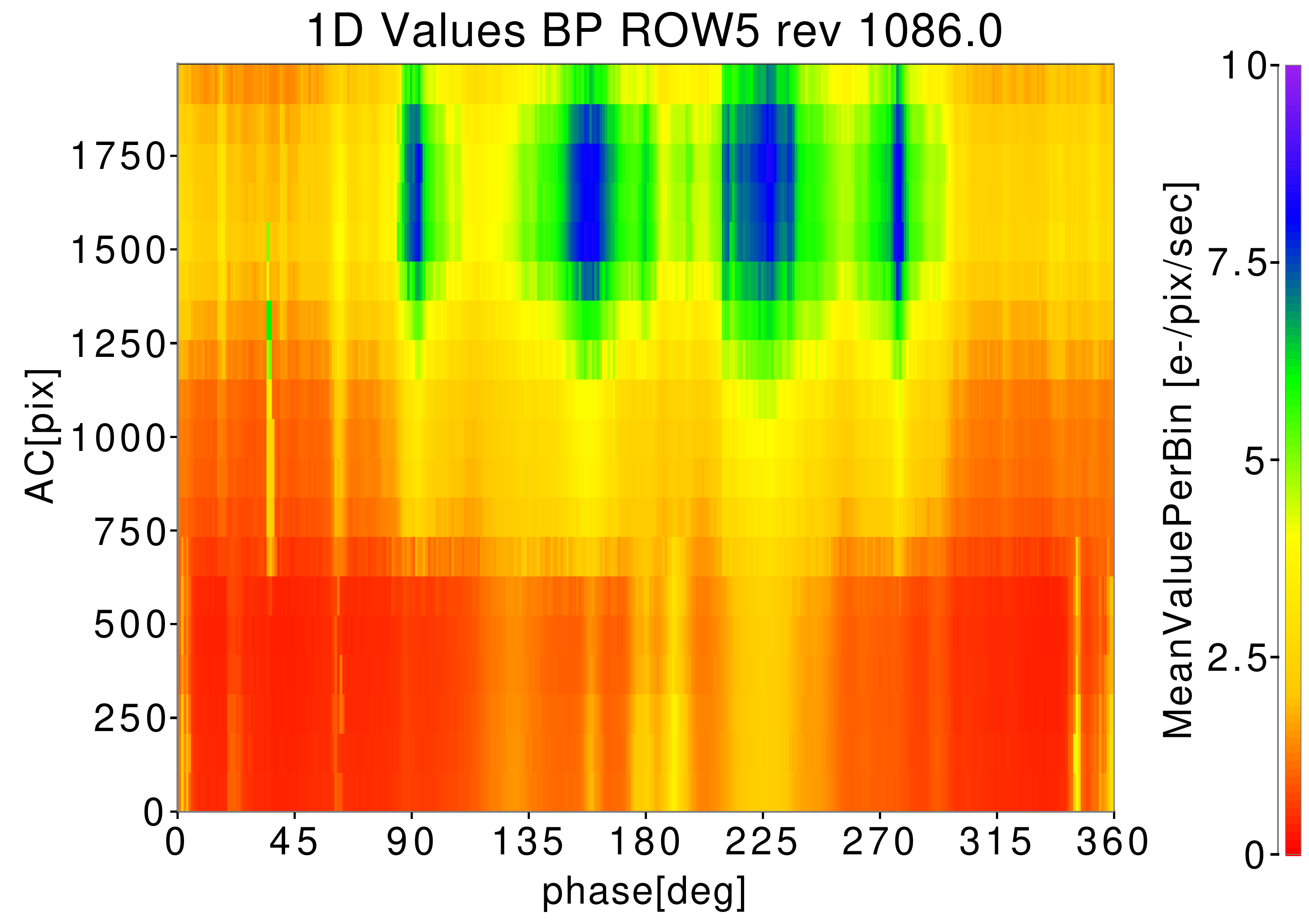}
\includegraphics[width=0.4\textwidth,angle=0]{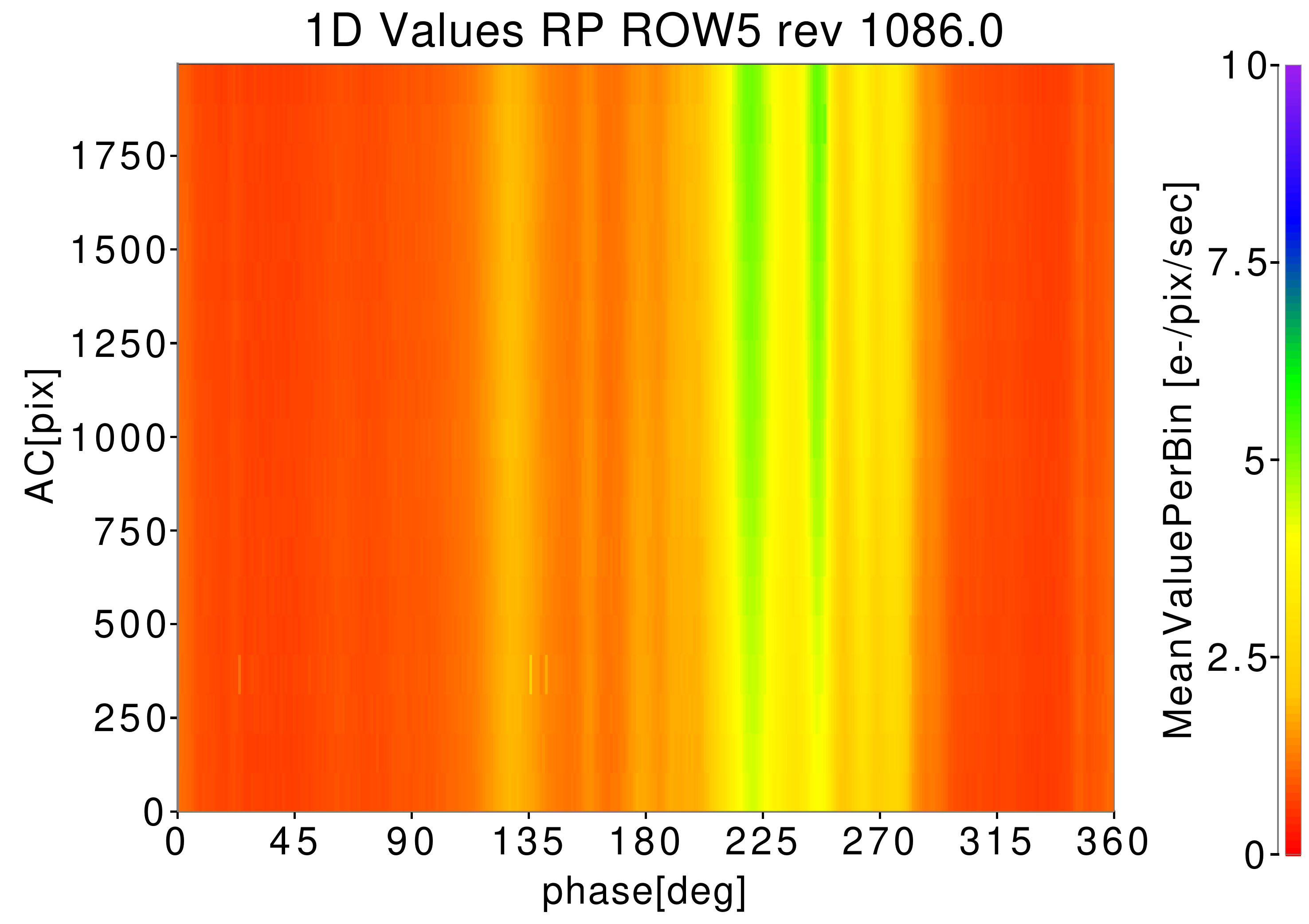}
        \caption{Stray light discrete maps for BP  and RP in the left and right plots, respectively,
         built with 1D observations (see text for details). In abscissa is the spin phase and in ordinate the AC coordinate. 
        The bin values in electron/pixel/s are colour-coded as in the bar to the left. Rev time in the figure label 
        indicates the start of the time range when the data were acquired, in this case revolution 1086.}
\label{Fig:slmapsOneD}
\end{figure*}

\begin{figure*}
\centering
\includegraphics[width=0.4\textwidth,angle=0]{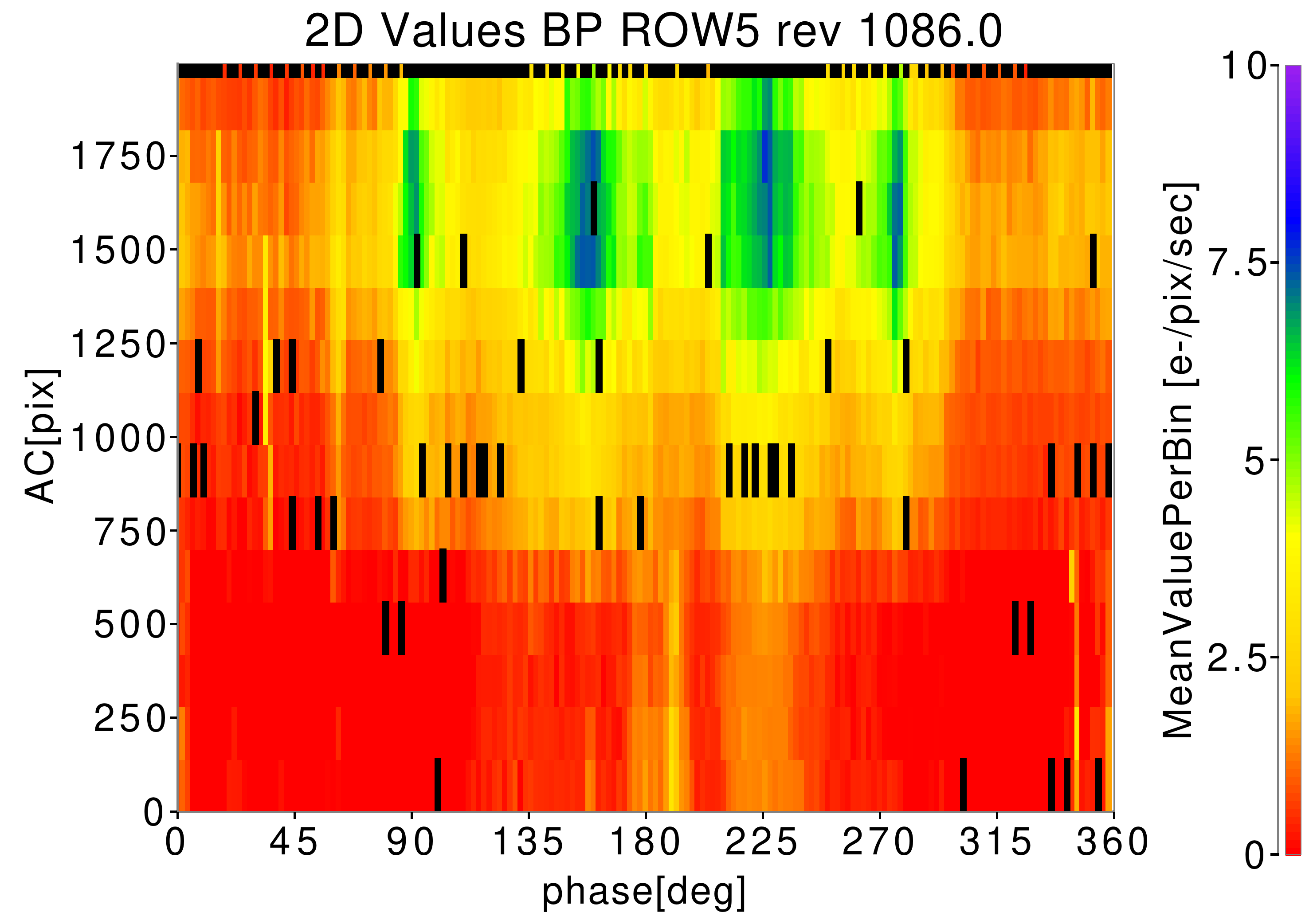} 
\includegraphics[width=0.4\textwidth,angle=0]{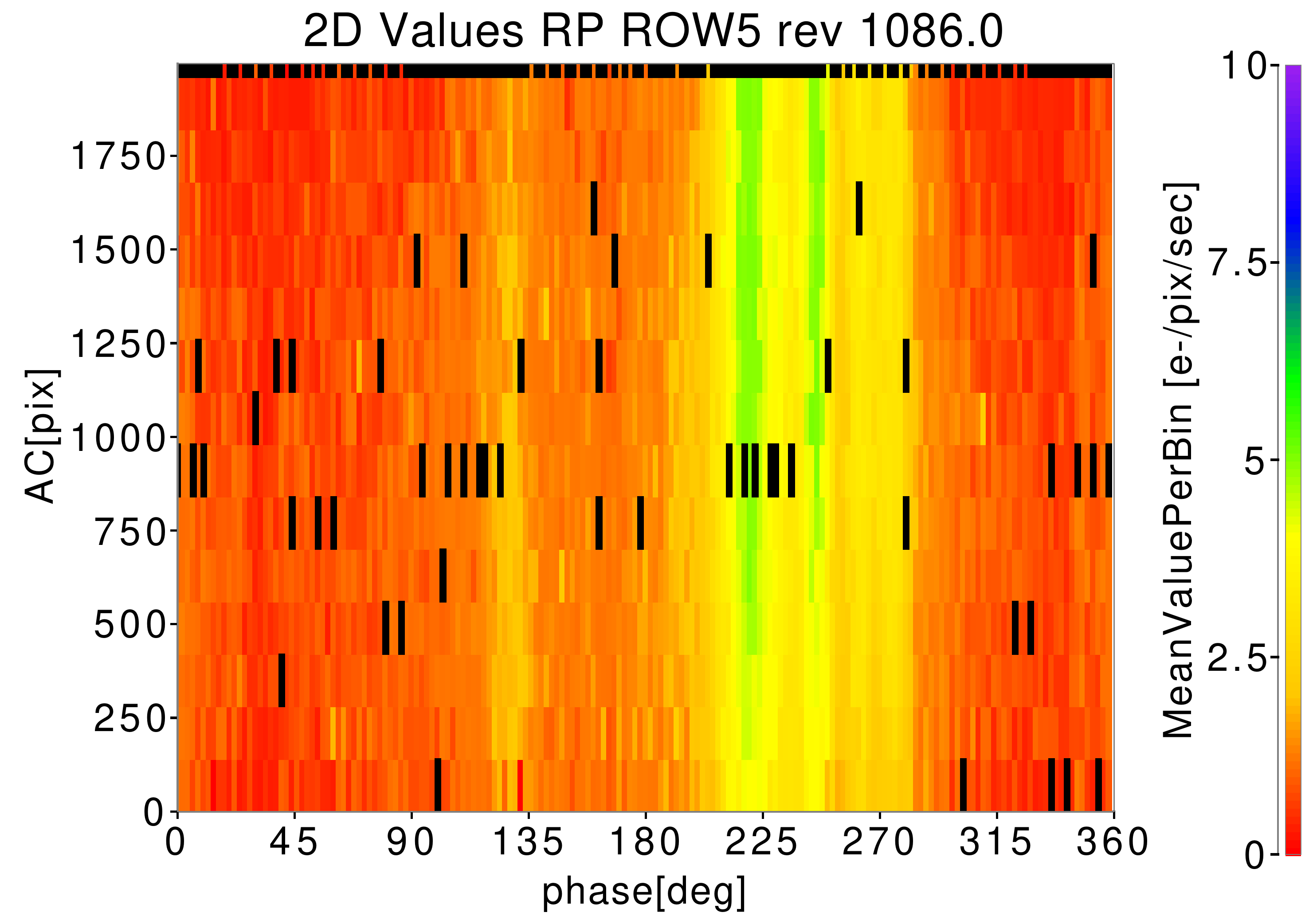}
        \caption{Same as in Fig. \ref{Fig:slmapsOneD} for 2D observations. It should be noted that the black squares indicate that there is  no information 
        for that bin. Interpolation is done for these cases. See \citet{PhotProcessing} for more details.}
\label{Fig:slmapsTwoD}
\end{figure*}

Examples of stray light maps obtained with 1D transits for BP and  RP are shown in Fig. \ref{Fig:slmapsOneD}, {while} maps obtained with 2D transits are shown in Fig. \ref{Fig:slmapsTwoD}.

\begin{figure}
\resizebox{\hsize}{!}{
\includegraphics[width=0.5\textwidth,angle=0]{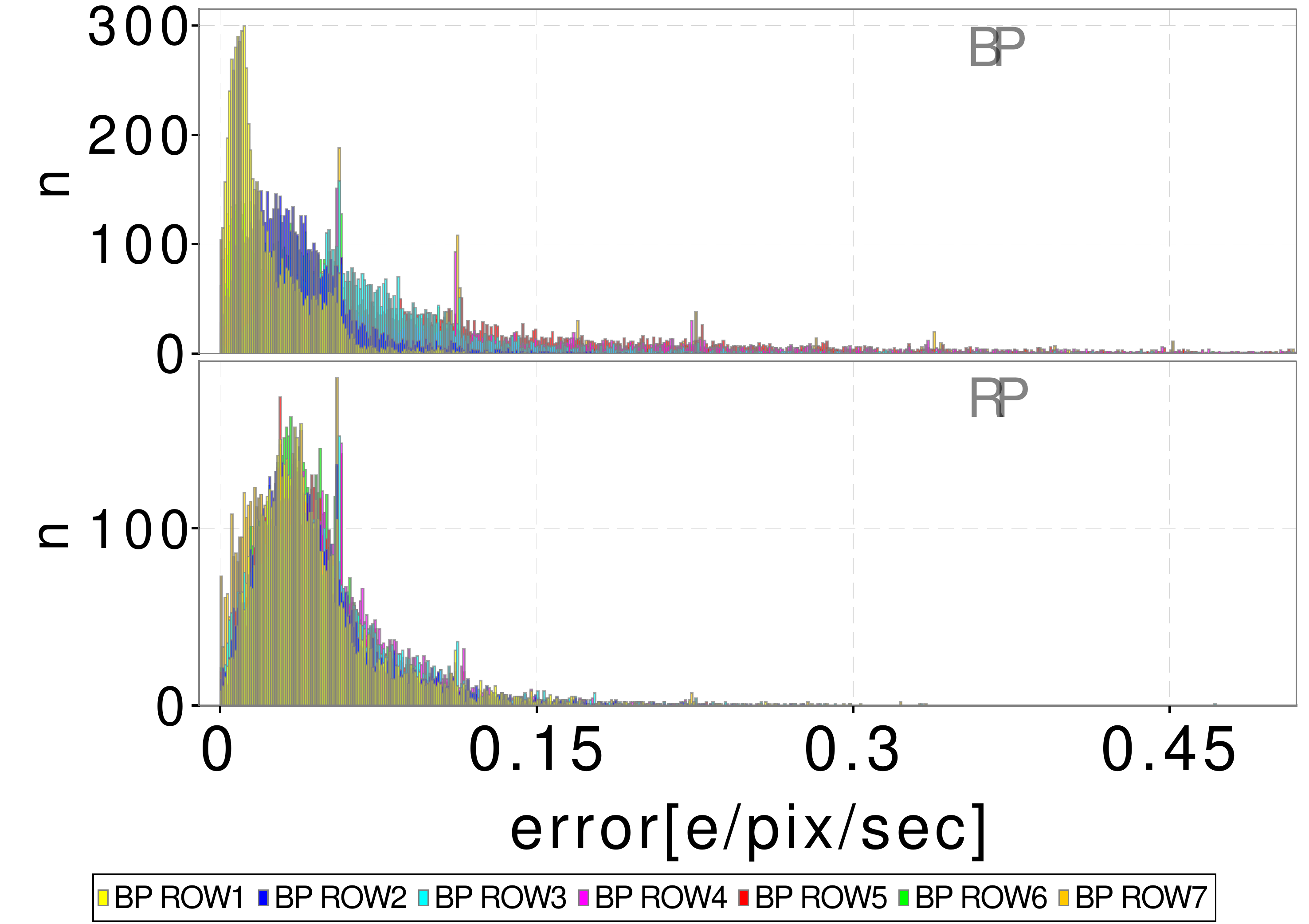}}
        \caption{Error distribution (in electron/pixel/s)  for the stray light maps built with 1D observations  (colour-coded by CCD row). 
        Top: BP. Bottom: RP.}
\label{Fig:slerrorsOneD}
\end{figure}

\begin{figure*}
\centering
\includegraphics[width=0.4\textwidth,angle=0]{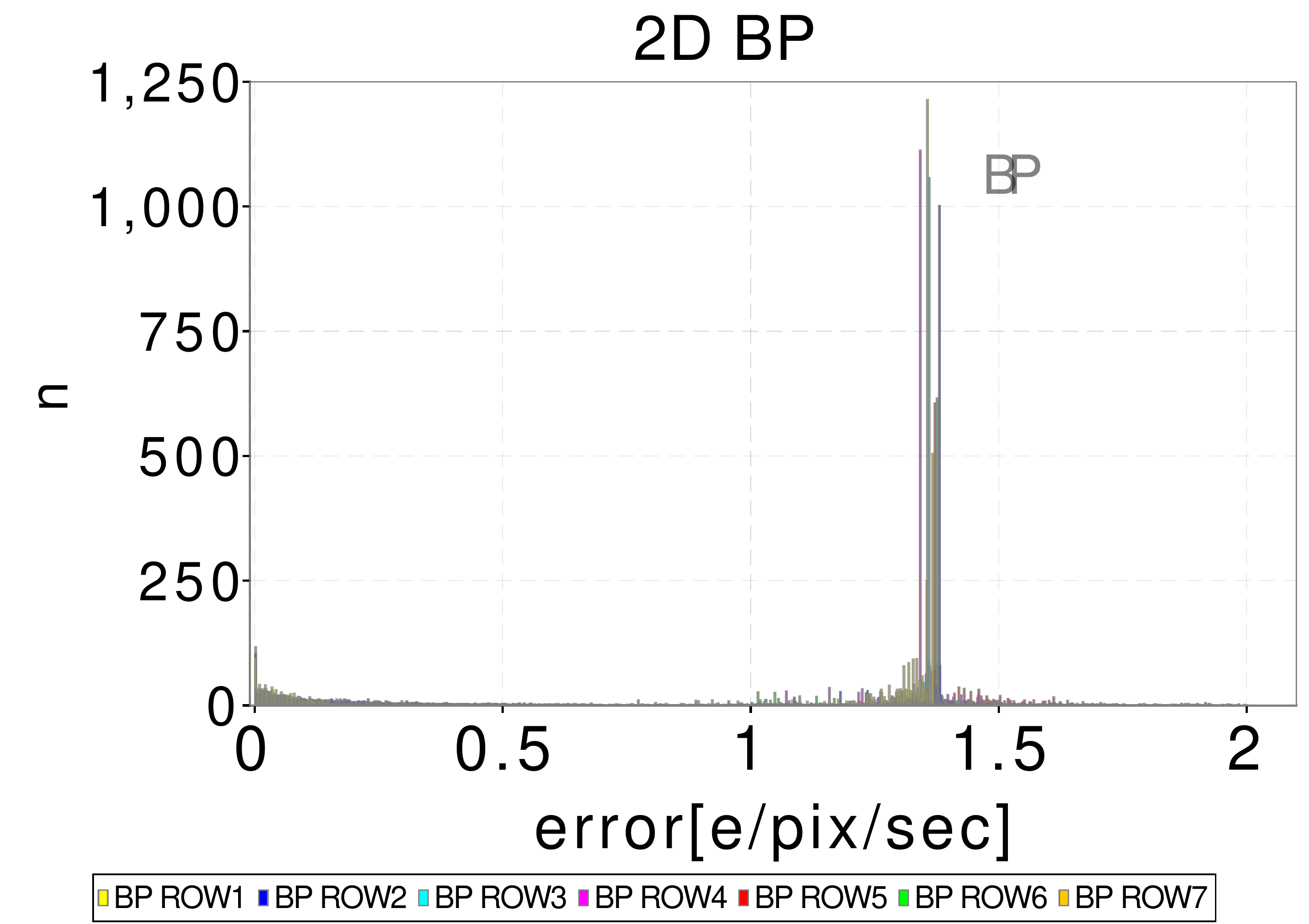}
\includegraphics[width=0.4\textwidth,angle=0]{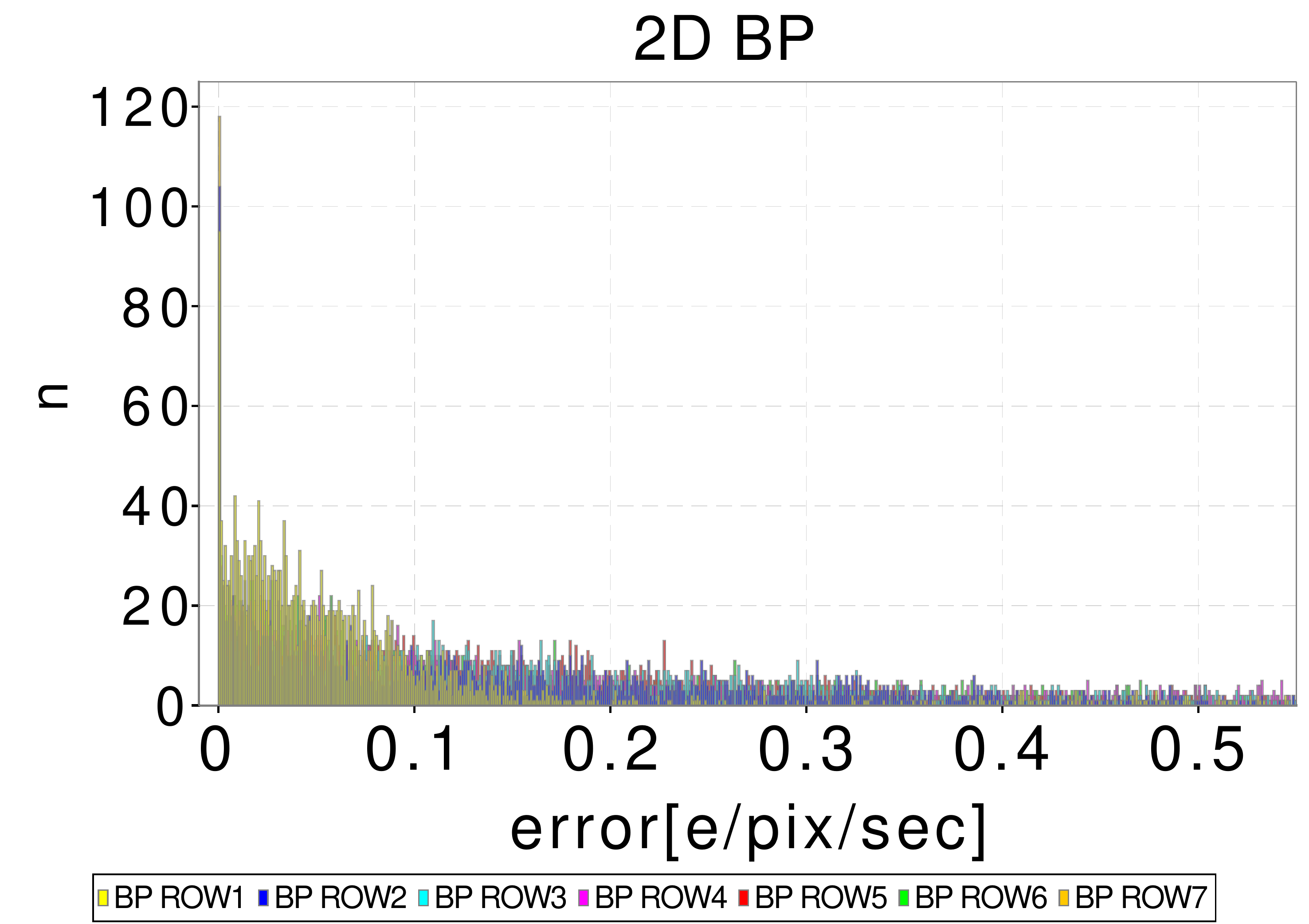}
        \caption{Error distribution (in electron/pixel/s)  for the stray light maps built with 2D observations for BP (colour-coded by CCD row).  The plot on the left shows all data, while the one on the right contains only the error values for the map bins with more than one measurement. See text for details.} 
\label{Fig:slerrorsTwoD_BP}
\end{figure*}

\begin{figure*}
\centering
\includegraphics[width=0.4\textwidth,angle=0]{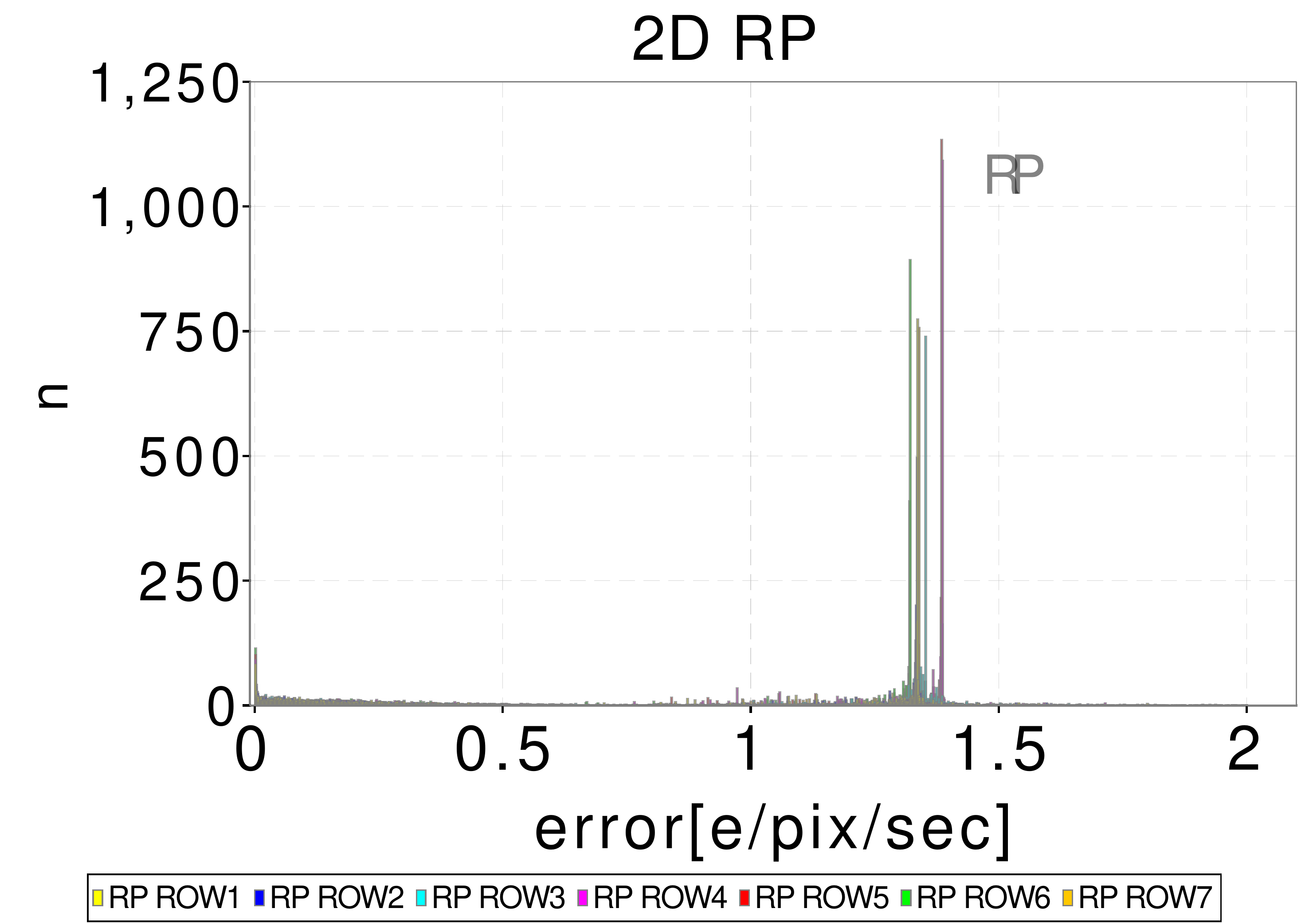}
\includegraphics[width=0.4\textwidth,angle=0]{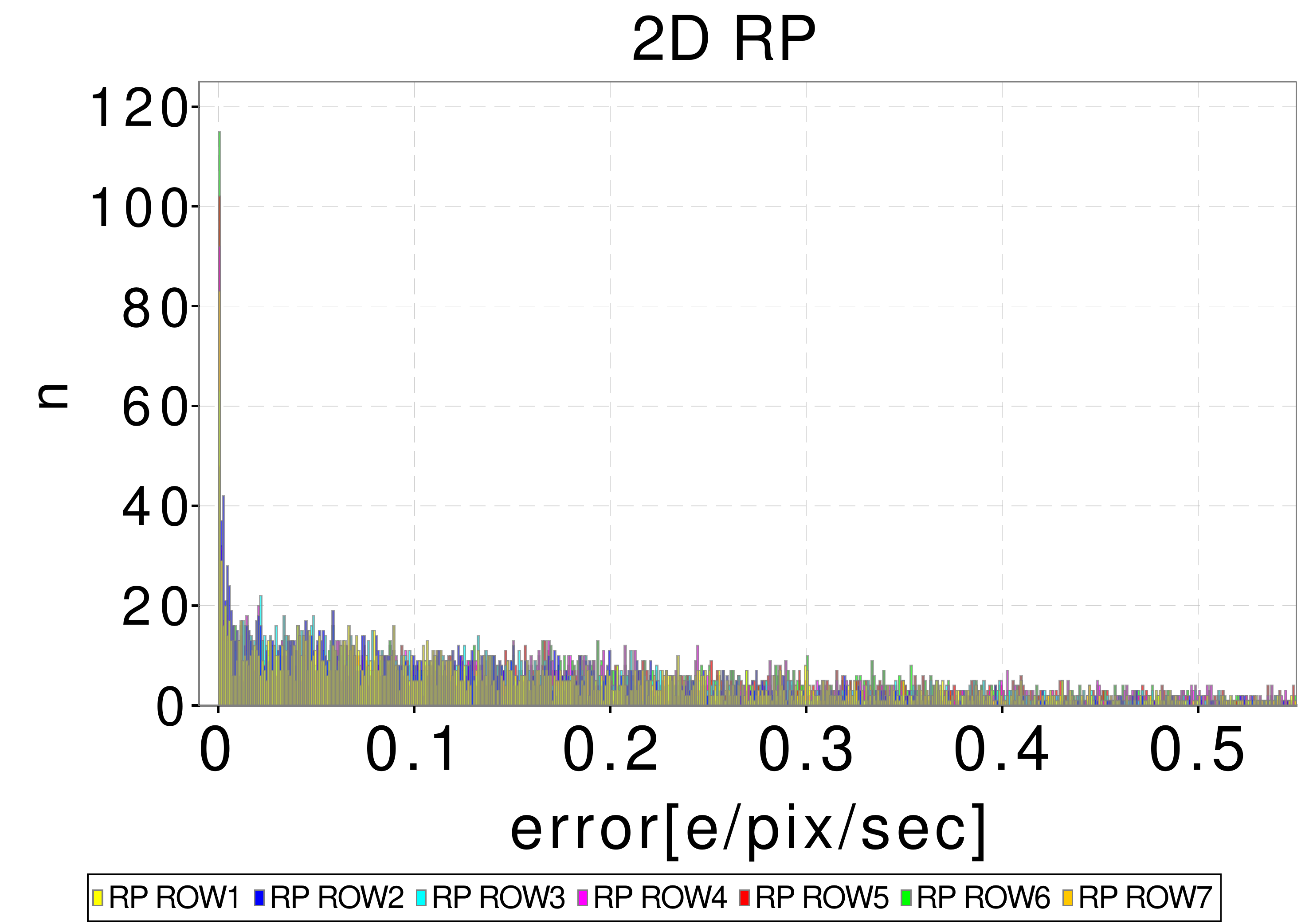}
        \caption{As in Fig. \ref{Fig:slerrorsTwoD_BP} for RP.} 
\label{Fig:slerrorsTwoD_RP}
\end{figure*}

A first validation is done directly by inspecting the map and looking at the distribution of the errors. The value in a bin is the median value obtained from all observations contributing to the bin. The median was chosen instead of the mean to reduce the effect of outliers caused for instance by cosmic rays or contamination from stars.
The error for a bin is calculated as the median absolute deviation (MAD) associated with the median value. 
Figures \ref{Fig:slerrorsOneD}, \ref{Fig:slerrorsTwoD_BP}, and \ref{Fig:slerrorsTwoD_RP} show the histograms for the stray light map bin errors, with a histogram bin size of 0.001 (electron/pixel/s). For the maps obtained with the 1D transits, the distributions are broadly similar for all CCDs.  
However, while in RP there is a close similarity between the rows, in BP there are more differences; this is explained by the fact that in RP the stray light features are similar for all CCDs, while in BP the features can change significantly in shape, position, and strength. 
For the maps obtained with the 2D transits, it is evident that there are two distributions: the first between 0 and 0.05, very similar to that for 1D transits and the second  between 1.35 and 1.4. The latter is due to the stray light bins with only one measurement, so that the error in that case is not the error on the median but the error on that single measurement. 

\begin{figure*}
\centering
\includegraphics[width=0.4\textwidth,angle=0]{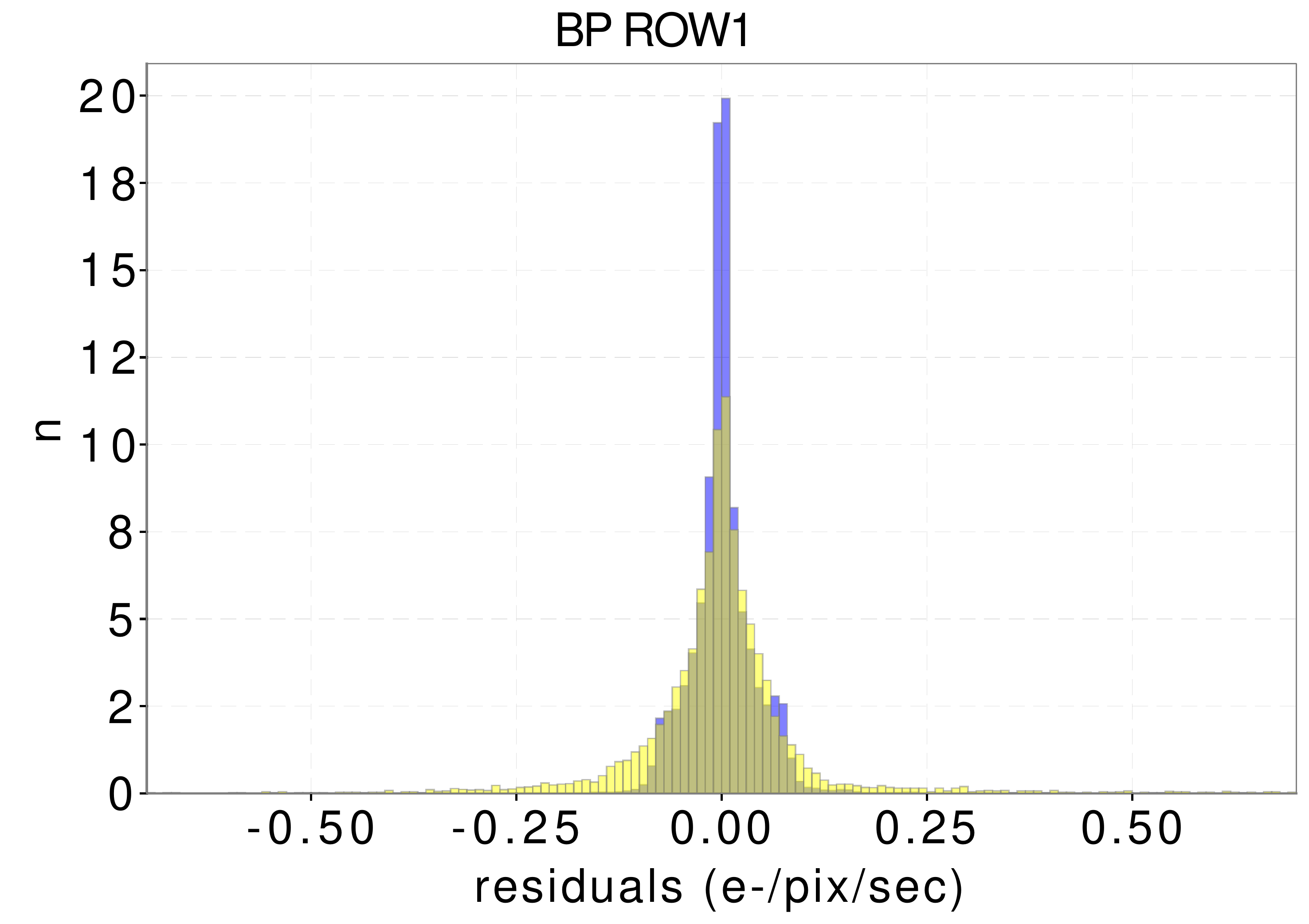} 
\includegraphics[width=0.4\textwidth,angle=0]{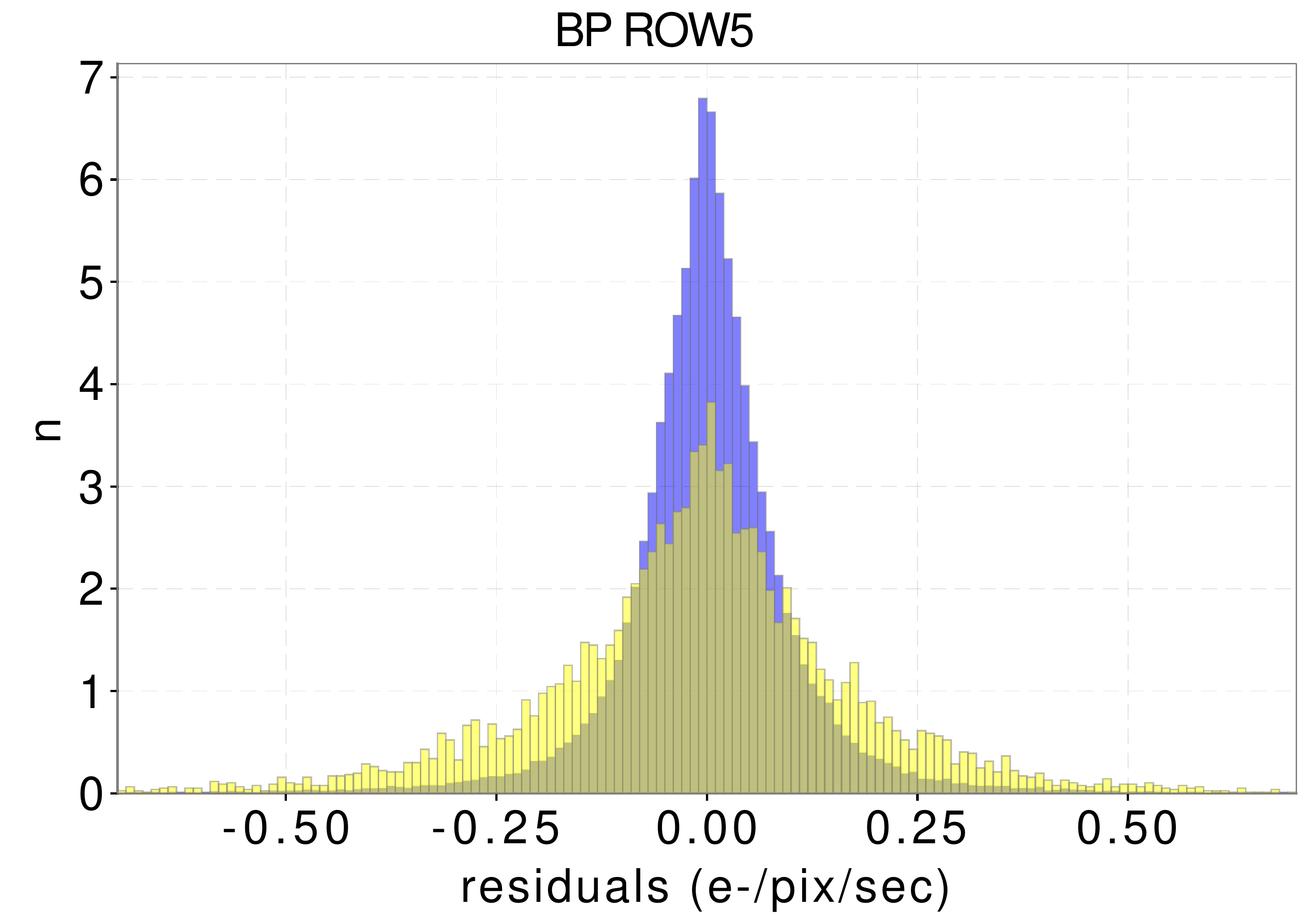}
        \caption{Residual distribution (normalised to the same area) for the stray light maps. In blue the data obtained from 1D transits, in yellow from 2D transits.
         The left plot shows the best case (for BP ROW1 with scatter $\sim 0.031$ for 1D data and $\sim 0.057$ for 2D data), while the right plot  shows the 
        worst case (for BP ROW5, with scatter $\sim 0.075$ for 1D data and $\sim 0.161$ for 2D data).
        }
\label{Fig:histo}
\end{figure*}

The residuals obtained by subtracting the map from the same data used to calculate it have also been analysed. 
 Figure~\ref{Fig:histo} shows examples of residual histograms. The histograms were normalised to the same area to allow a better comparison. The distribution is well centred around zero, with a different width for 1D and 2D transits, showing that the model is correct and that there is no residual trend.

An additional check is the calculation of the scatter of the residuals, made using an interquartile method used in the Hipparcos mission \citep{Hipparcos} which uses the percentile values at 15.85 and 84.15\% to robustly estimate the standard deviation of the distribution (see also Sect.~\ref{Sect:Constant}).  The scatter is lower for 1D than 2D observations, as shown
in Fig.~\ref{Fig:histo} (left panel) where the value is $\sim 0.053$ electron/pixel/s for residuals from 1D observations, while it is $\sim 0.131$ electron/pixel/s for residuals from 2D observations. This is expected, since for 2D windows the number of observations is much lower (about 10\% of the number of 1D windows) and therefore the model is less accurate and the residuals are bigger. The same results apply to RP as well.
The worst case, shown in { the right panel of} Fig. \ref{Fig:histo}, is when the variations in  the  AC and  AL directions are quite large, but  this is expected as well since the resolution of the map is not  sufficient to reproduce  the rapid variations.  Unfortunately, increasing the resolution of the map  is not an option because there is simply not enough data available to robustly measure the background  level. 
This scatter translates into an error in magnitude which is well below the expected end-of-mission error: in the worst case, the values are comparable but it should be noted that  the end-of-mission error is  calculated based on the accumulation of the measurements, while the scatter is calculated on single measurements and  will decrease.

\section{Study of LS and SS calibration coefficients}

As described in \citet{PhotPrinciples}, two of the main photometric calibrations are referred to as the large-scale (LS) and small-scale (SS) calibrations. They can be used  for validation in two ways.
The first is used in the validation of the calibrations themselves, and the second is used in the validation of the photometry as a whole in the detection of anomalies.

\begin{figure} \resizebox{\hsize}{!}{\includegraphics{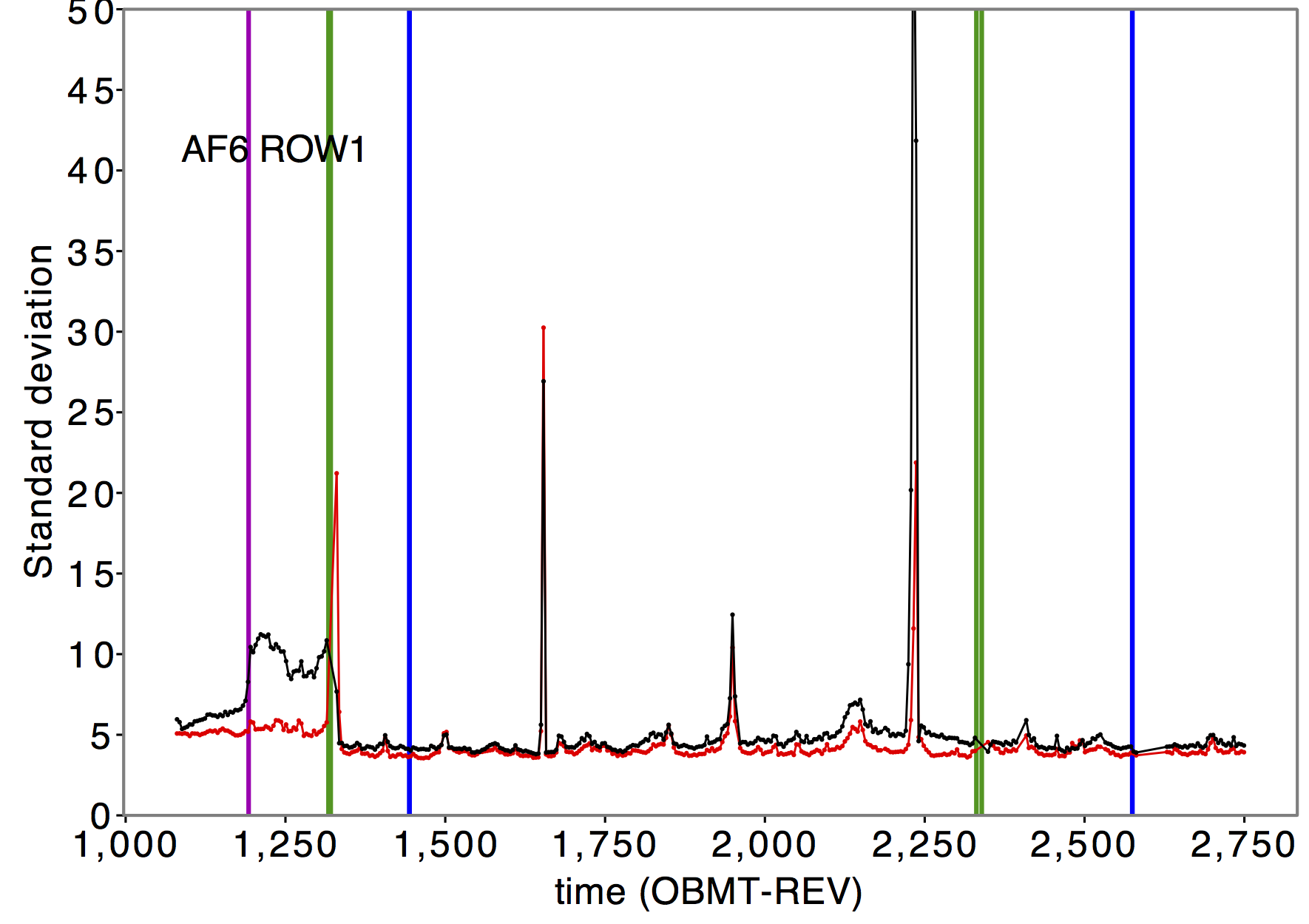}} \caption{Unit-weight standard deviation of the large-scale calibration as a function of
time (in satellite revolutions) for an example calibration unit. In this case, AF6, Row 1, Window Class 1, No Gate.
The black lines are for the preceding and red for the following FoV calibration units. The vertical lines represent significant satellite events: scanning law change (magenta), decontamination (green),
and refocussing  (blue).}
\label{Fig:SdLs}
\end{figure}

\begin{figure} \resizebox{\hsize}{!}{\includegraphics{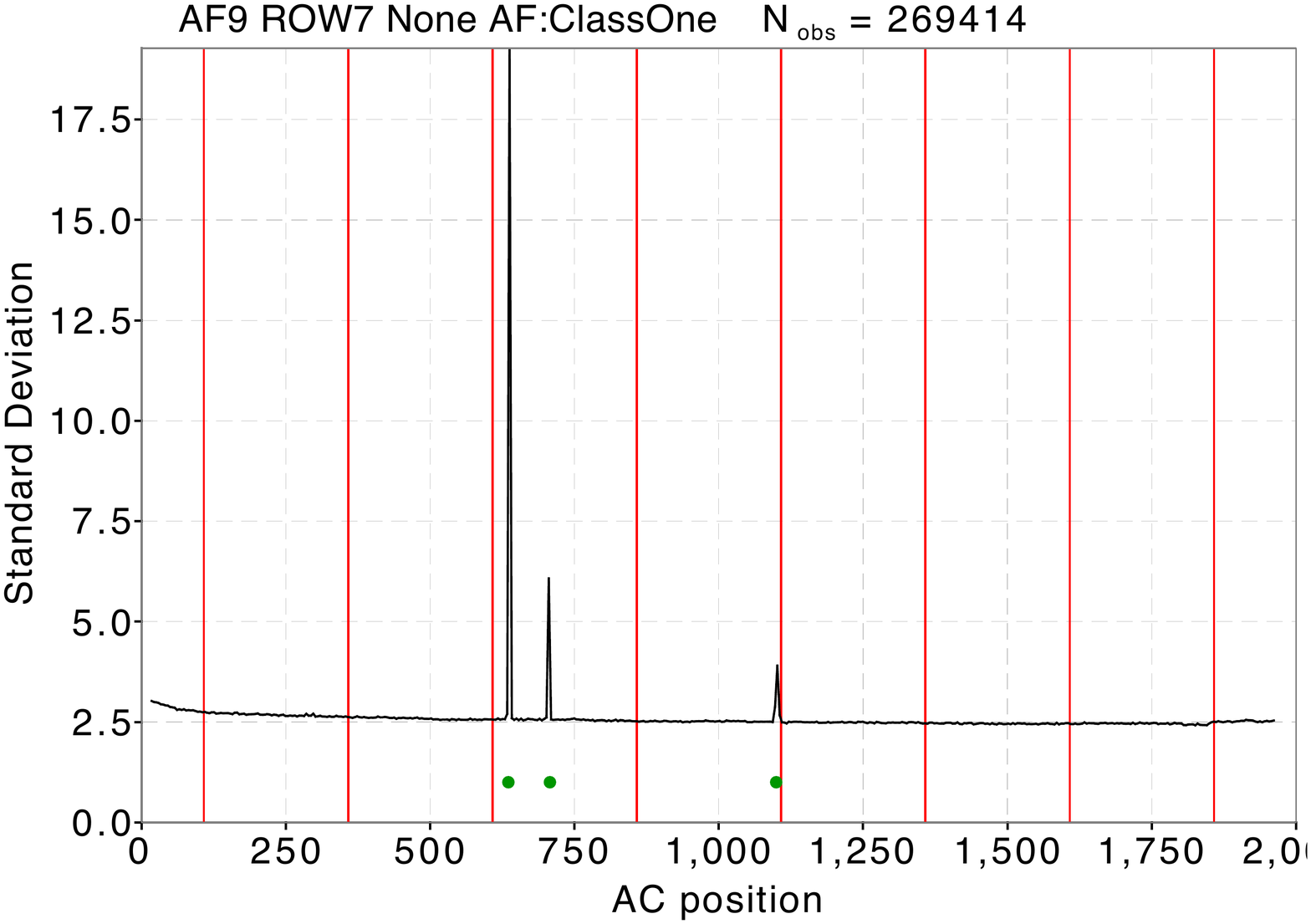}} \caption{Unit-weight standard deviation of the small-scale calibration as a function of
across-scan position on the CCD for an example calibration unit. In this case, AF9, Row 7, Window Class 1, No Gate. The red lines show the locations of the CCD stitch blocks and the green dots
show the location of detected  bad columns.}
\label{Fig:SdSs}
\end{figure}

When the calibrations are carried out, the unit-weight standard deviation of the solution is calculated. 
 This is defined as the square root of the normalised chi-square \citep{NewHipparcos}.
This gives an indication of how well the solution model is able to remove
any systematic effects. In the ideal case, this value should be around 1.0. However, in these early stages of the mission, it is not expected that the values found in the solutions
would be  close to ideal,   either because  the calibration model does not represent the systematics very well or because the quoted errors on the fluxes do not 
correctly represent the true error (or both). 
Figures \ref{Fig:SdLs} and \ref{Fig:SdSs} show example plots of the standard deviation for the large- and small-scale calibrations, respectively.
Where the standard deviation varies from the average value, it indicates a region where the calibration model is worse at modelling the systematic effects 
 and that a possible improvement or additional calibration feature is required. 
In the example of the large-scale calibration (Fig.~\ref{Fig:SdLs}),
the average value of 5.0 implies that the observed scatter in the data for this configuration will be 5 times worse than the quoted errors for those periods with a
standard deviation of 5.0. This only affects the individual transit measurements. It should be noted that the error on the weighted mean will not be affected by this since the 
measured scatter has been accounted for in its calculation \citep[see][for more details]{PhotPrinciples}.

In the example shown for the large-scale calibration (Fig.~\ref{Fig:SdLs}), the peaks seen correspond to short periods, sometimes individual calibrations, which indicate problems with the IPD 
\citep[see][for more details]{IdtRef}, such as the use of an incorrect or suboptimal LSF/PSF library. Future processing cycles will use redetermined IPD values  
for which many of these features will have been corrected. The period immediately after the first
decontamination (within the period covered by \GDR1) may be problematic owing to the focal plane possibly not having reached thermal stability. The quality of these few days is being  investigated further.
Also seen in this plot is an indication that the period between the change in the scanning law and the first decontamination is of a poorer quality for the preceding FoV in comparison to the rest of the
\GDR1 period. It should be noted that for \GDR1, the calibrations are  carried out approximately every day, which is how time variation in the response function is calibrated. 

For  the  small-scale calibration (Fig.~\ref{Fig:SdSs}), the main features seen in the standard deviation plots arise from bad columns. Many of these are
confirmed in the CCD health calibrations  \citep[see][for more details]{IdtRef}. In the later stages of the mission,  this information will be used to mask the affected samples as part of the PSF fit of 2D windows performed by the IPD process \citep[see][]{IdtRef}.

\begin{figure} \resizebox{\hsize}{!}{\includegraphics{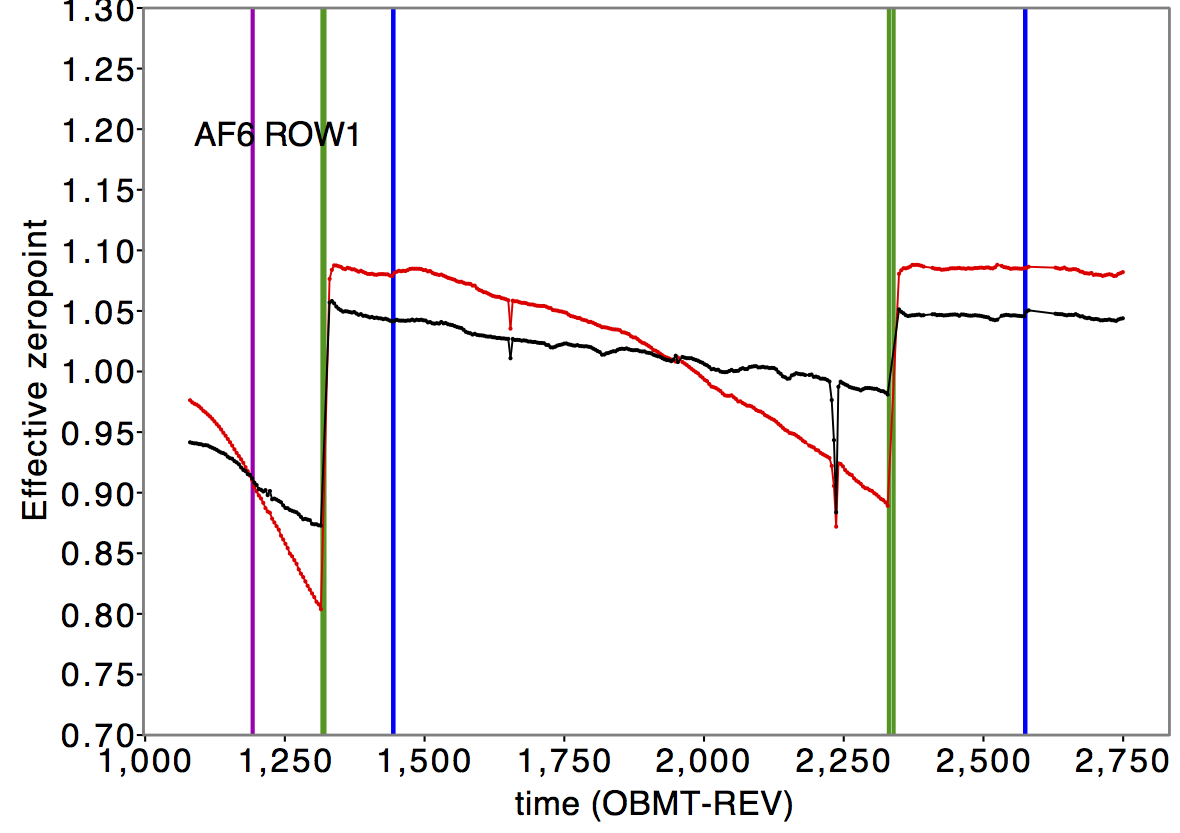}} \caption{Effective zeropoint of the large-scale sensitivity calibration as a function of
time for an example calibration unit. The calibration unit is the same as in Fig.~\ref{Fig:SdLs}, as are the vertical lines.
For this plot the SSC terms of the calibration model have been combined to form an effective zeropoint  using default colours.
This is necessary since there is no zeropoint term in the calibration model.
See \citet{PhotPrinciples} for more details.}
\label{Fig:ZpLs}
\end{figure}

\begin{figure} \resizebox{\hsize}{!}{\includegraphics{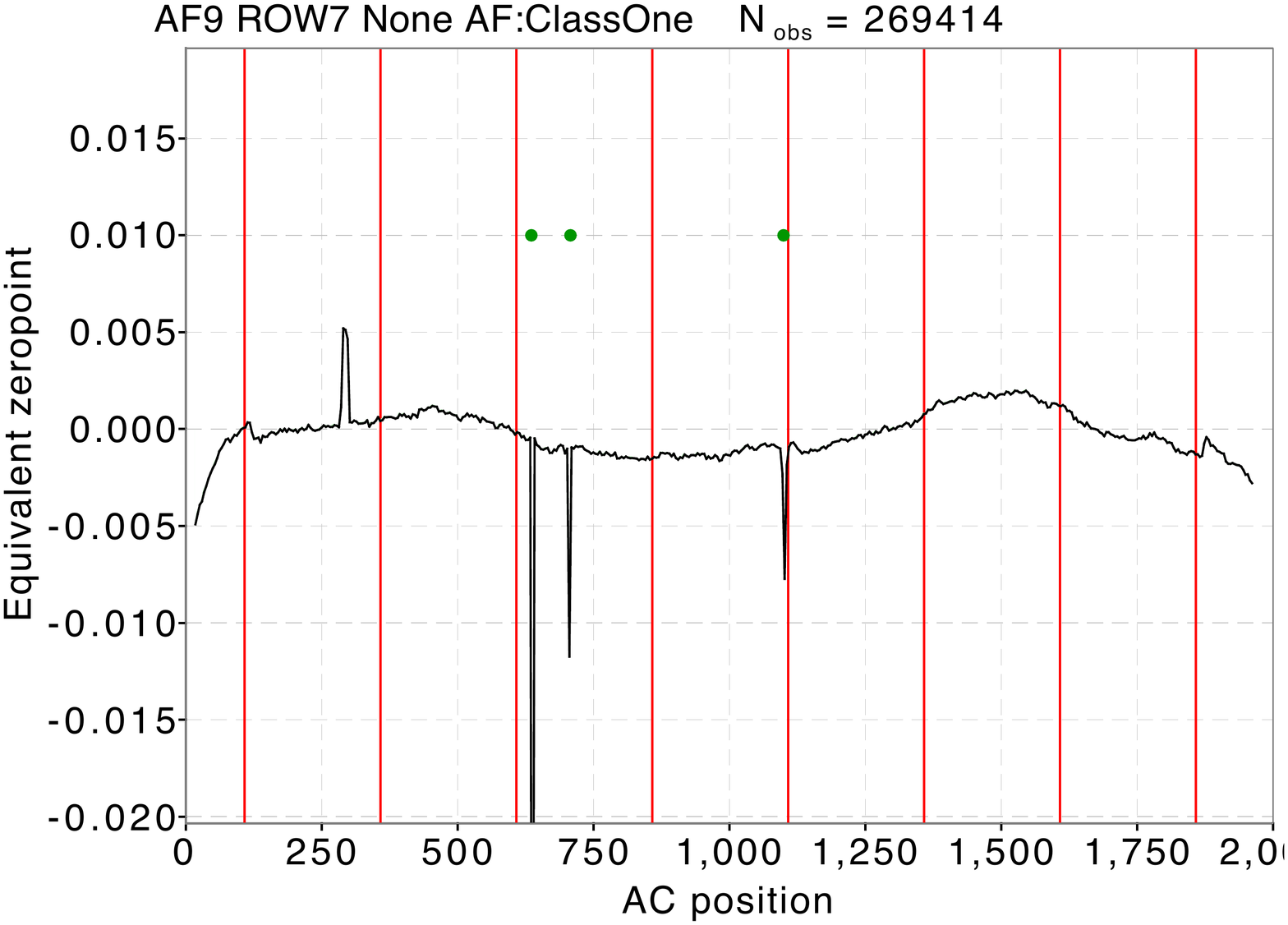}} \caption{Zeropoint of the small-scale sensitivity calibration as a function of
across-scan position on the CCD for an example calibration unit. As defined in \citet{PhotPrinciples}, 1.0 has been subtracted from the zeropoint.
The calibration unit is the same as in Fig.~\ref{Fig:SdSs}. The red lines show the locations of the CCD stitch blocks and the green dots
show the location of detected  bad columns.}
\label{Fig:ZpSs}
\end{figure}

Plotting the various calibration coefficients from the solutions as a function of time (LS) and AC position (SS) is also a good way to identify anomalies and to indicate where
further investigation is required (see Figs. \ref{Fig:ZpLs} and \ref{Fig:ZpSs}).

The  main features seen in the large-scale calibration plots (Fig.~\ref{Fig:ZpLs}) are the changes in the response of the CCD due to the varying levels of contamination on the mirrors and CCDs. As the
mission progressed, more contaminant was deposited on the mirrors and CCDs, thus reducing the efficiency of the overall system. The response is different between the two  FoVs;  the
mirrors associated with the following FoV are more highly contaminated.  As already mentioned in Sect. \ref{Sect:BpRpCal}, two decontamination campaigns were performed during the period covered by \GDR1. This successfully improved the photometric throughput as  can be seen from Fig.~\ref{Fig:ZpLs}. However, the contamination  was not fully removed and continued to increase with time, albeit at a reduced rate.
It should be noted that some of the spikes seen in the standard deviation plots (Fig.~\ref{Fig:SdLs}) are also seen in the coefficient plots (Fig.~\ref{Fig:ZpLs}). 

The main variation mapped out by the small-scale calibration is the response as a function of  AC position on the CCD (Fig.~\ref{Fig:ZpSs}). This is effectively a 1D flat field. 
Again, the bad columns present on the \Gaia\ CCDs can be seen in the zeropoints  of the small-scale calibrations. When combined with the matching standard deviation plot
(Fig.~\ref{Fig:SdSs}), this shows that the current model is not appropriate for these columns.  This plot  shows also  a small variation in the response { at around AC position 300} for 
a small number of columns. In this case, there is no corresponding spike in the standard deviation plot indicating that the model is reasonably correct and that this does represent a genuine response variation.

\begin{figure} \resizebox{\hsize}{!}{\includegraphics{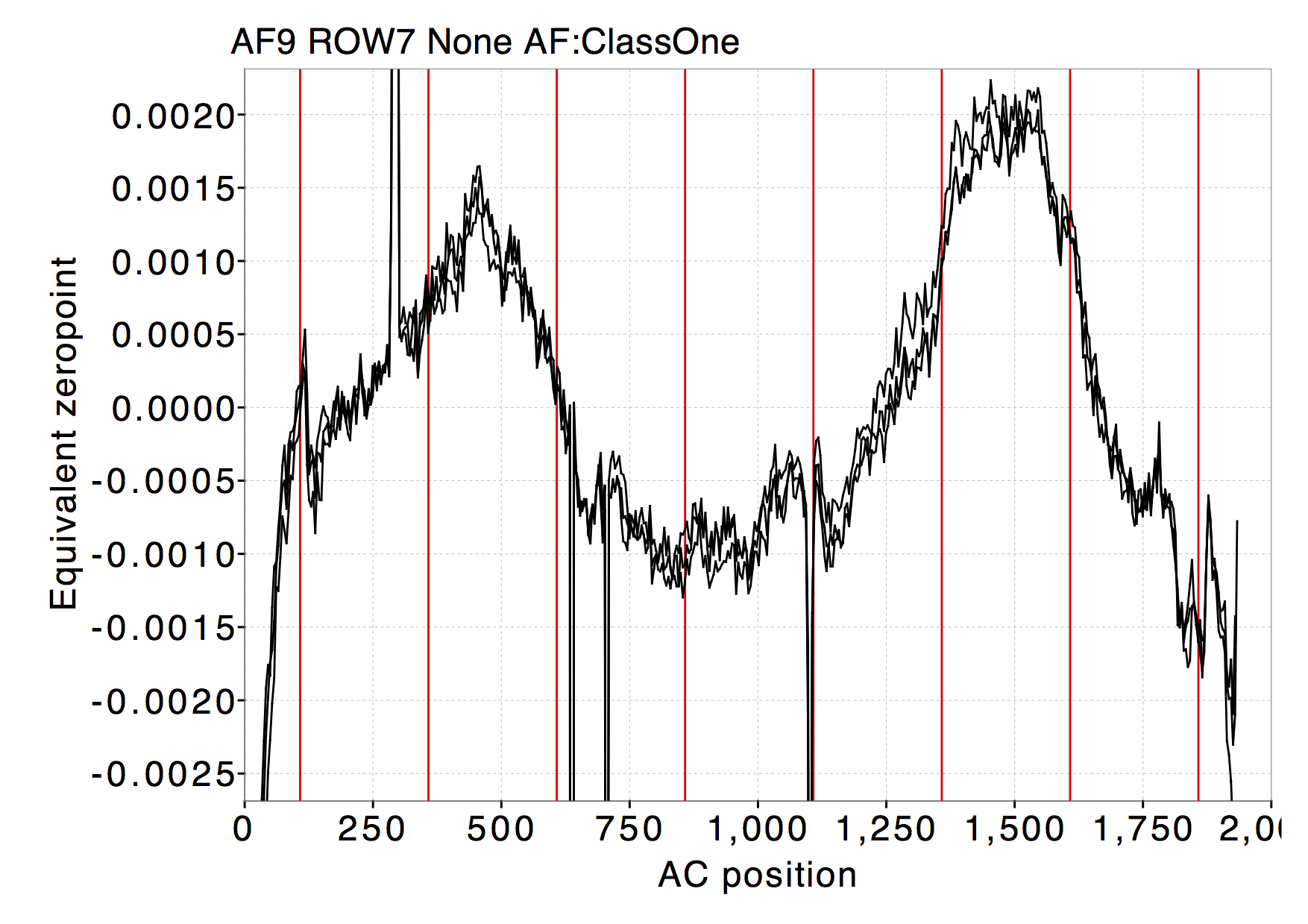}} \caption{ Zeropoint of the small-scale sensitivity calibrations as a function of
across-scan position on the CCD (same calibration configuration as Fig.~\ref{Fig:SdSs}) where the time range of \GDR1 has been divided giving three sets of calibrations. We note the change in
the ordinate scale. This plot was derived from a preparatory processing run (OR5S3).}
\label{Fig:SsVar}
\end{figure}

It should be noted that for \GDR1,  only a single set of SS calibrations spanning the entire time range was computed in order to ensure enough calibrators at the bright end of the magnitude scale.  In order to verify that the SS calibrations are indeed stable over the entire time range, the period was divided into three and a set of calibrations was derived for each one.
No significant variation was seen between the three { sets of} calibrations at the level that was required for \GDR1 (see Fig.~\ref{Fig:SsVar}). The variation in this plot, 
typical for Window Class 1 (13$<G<$16), 
was 0.17 mmag
as measured by the robust width mentioned earlier.

\section{Convergence of the large-scale calibrations}\label{Sect:Convergence}

\begin{figure} \resizebox{\hsize}{!}{\includegraphics{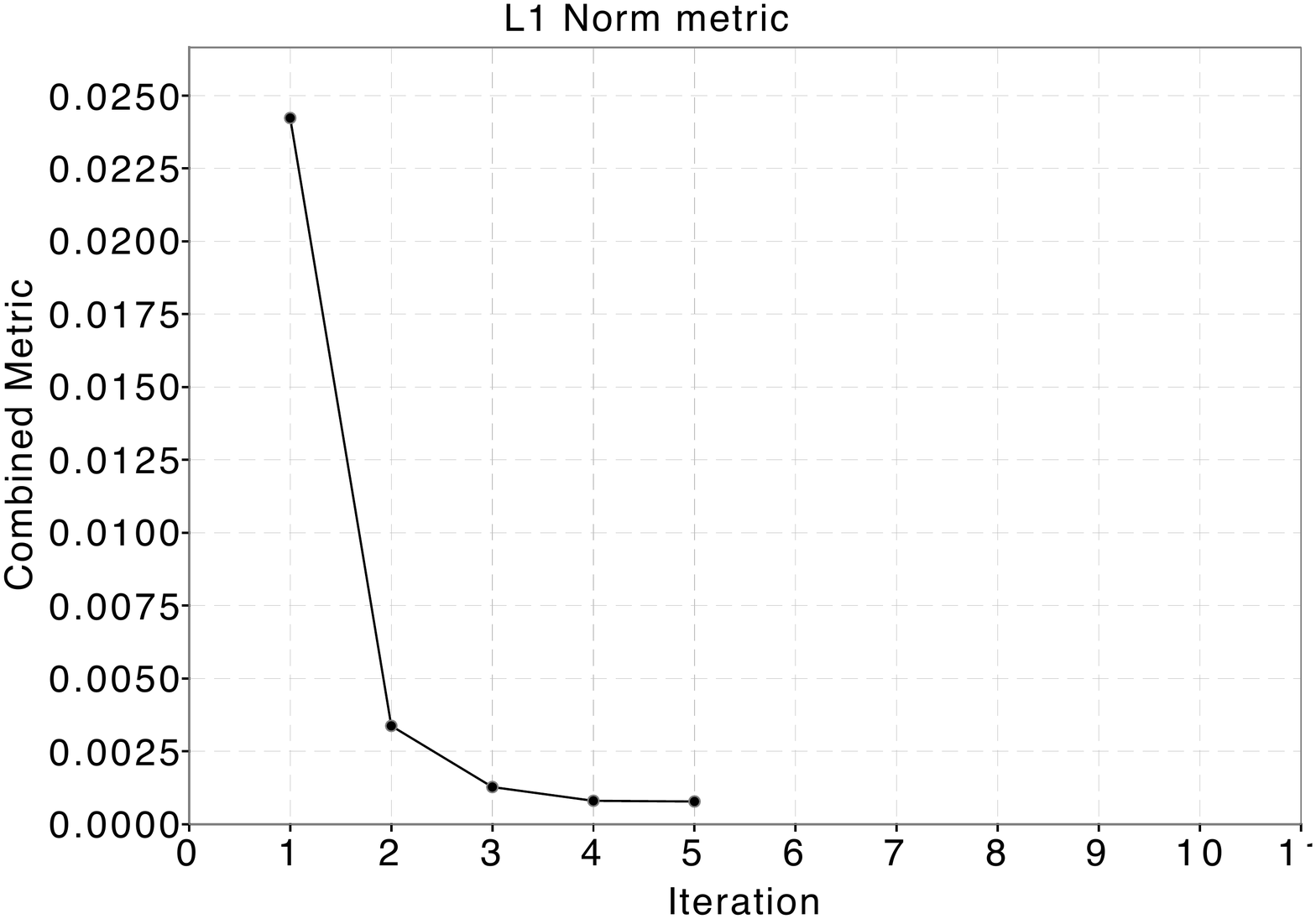}} \caption{Convergence metric as a function of iteration for the G-band Window Class 2 large-scale
calibrations. These metrics compare the large-scale
calibrations between two iterations, the one numbered in the plot and the following one. The exception is the final point which compares the large-scale calibrations at the end of the initial set of iterations
and those done after the small-scale calibrations have been carried out; see \citet{PhotPrinciples} and \citet{PhotProcessing} for more details.}
\label{Fig:Convergence}
\end{figure}

As described in \citet{PhotPrinciples},  the photometric system needs to be established in the initial stages of calibration. This is done by iterating between the large-scale calibration and determining 
the reference fluxes using the latest iteration of calibrations. In order to show that the system is converging, a form of convergence metric needs to be used. The one chosen was an 
L1 norm and was chosen in preference to the L2 norm since it is more robust to outliers. The general form of the L1 norm is
\begin{equation}
\int |p_{i}({\bf x})-p_{i+1}({\bf x})|d{\bf x}
,\end{equation}
where $p_{i}$ corresponds to the calibration factor for the $i$th iteration and ${\bf x}$ the singular parameters of a source. If this integral is carried out over a representative range of
parameter space, the norm represents the typical change in the calibration factors when going from one iteration to the next. In this analysis, this was done by using the singular parameters (e.g.~colour) of about 1000 randomly selected sources.
The overall metric used was the median value of the norms for the calibrations considered.

Figure~\ref{Fig:Convergence} shows the convergence metric for the G-band Window Class 2 calibrations ($G>16$). The final data point shows the difference between the final LS calibrations of the
iteration stage and the LS calibrations performed after the SS calibrations have been carried out.
It can be seen that the photometric system converges very well.
After five iterations were carried out, it was decided to stop the initialisation process considering that the changes had reached the mmag level. In future releases,  
further iterations will be carried out to improve on this performance.

\begin{figure} \resizebox{\hsize}{!}{\includegraphics{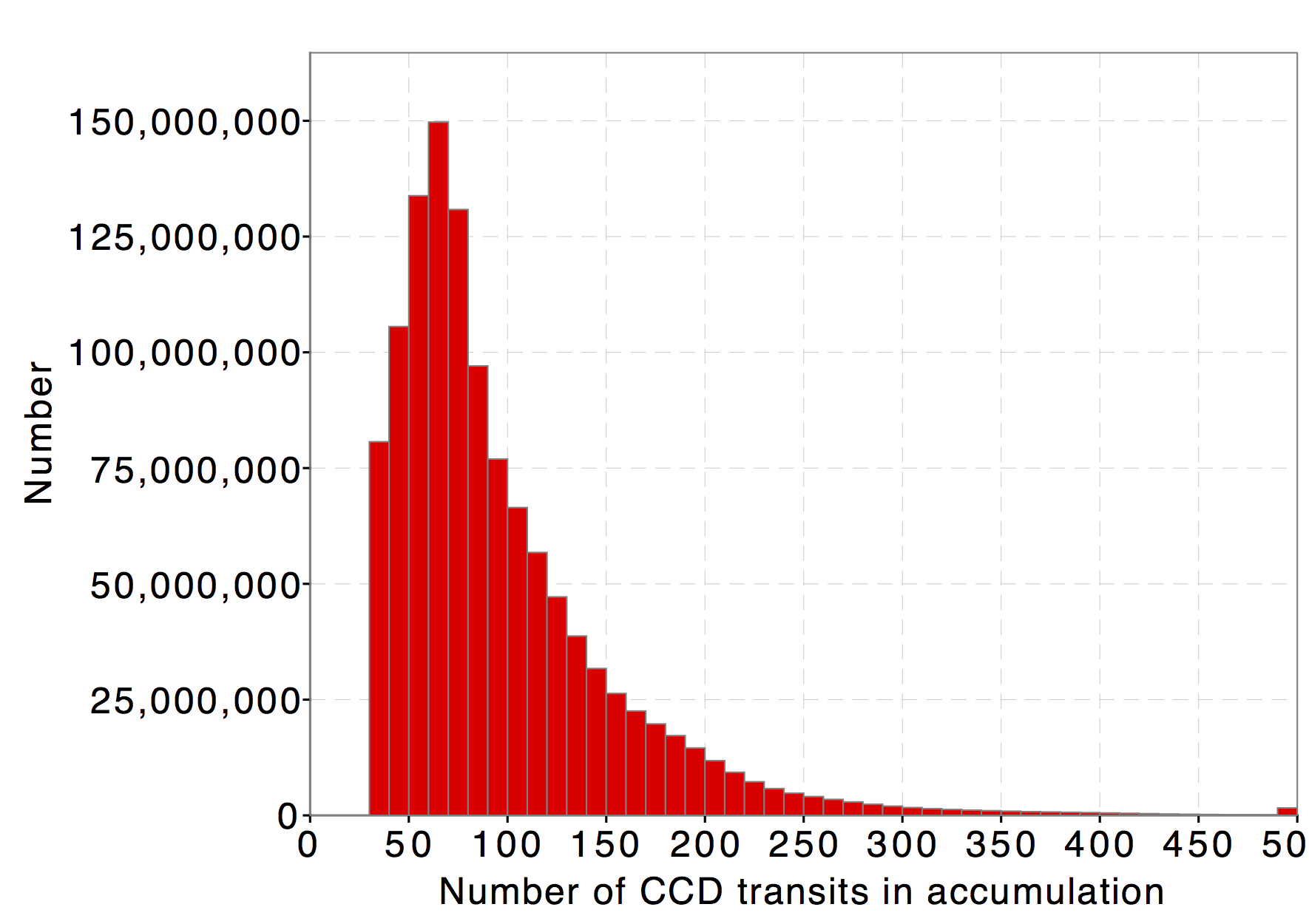}} \caption{Distribution of the number of G-band CCD transits for each source analysed.}
\label{Fig:AccumNobs}
\end{figure}

\section{Analysis of accumulation data}\label{Sect:Accumulation}
As described in \citet{PhotPrinciples}, the data for each source is accumulated and various statistics gathered. Figure~\ref{Fig:AccumNobs} shows the distribution of the number of 
G-band CCD transits for each source analysed. { T}o remove most of the spurious detections
made by \Gaia, which are mainly around bright sources{, the validation analysis has a lower cut-off of 30 CCD transits (roughly corresponding to 3 FoV transits) as seen in this histogram.} Because they are  spurious, such detections are unlikely to be matched with other observations \citep[see the section on cross-matches  in][]{IdtRef} and such ``sources'' will thus have low numbers of CCD transits accumulated.
The average number of G-band CCD transits for \GDR1 is just under 100 (with mean and median equal to 97 and 79, respectively) which corresponds to about 10 FoV transits. The spread in the number of observations is due to the
scanning law, and some sources will have significantly more observations than  the average.

\begin{figure} \resizebox{\hsize}{!}{\includegraphics{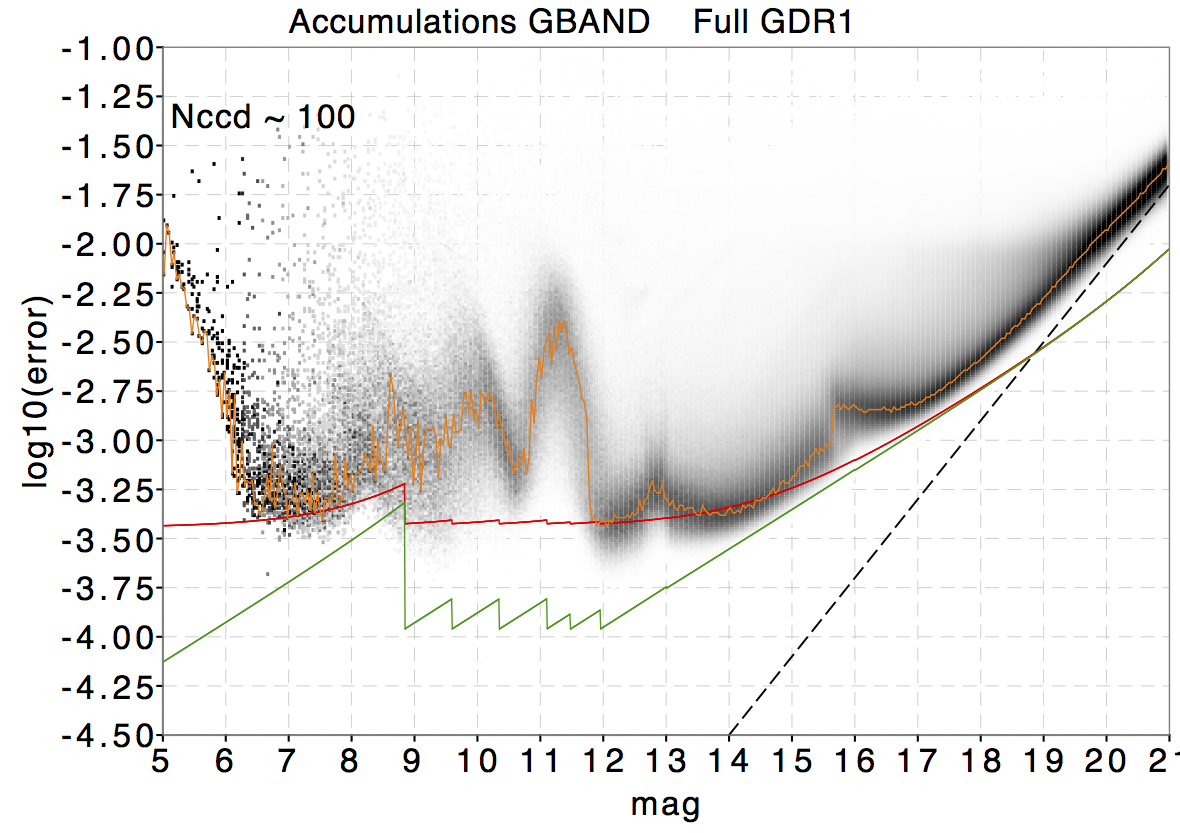}} \caption{ Distribution of error on the weighted mean G-value as a function of magnitude. The orange line shows the mode of
the distribution.
This plot is restricted to all sources with between
90 and 110 CCD transits. The green line shows the expected errors for sources with 100 CCD transits and for a nominal mission with perfect calibrations. 
The red line shows the same error function, but with a calibration error of
3 mmag added in quadrature to the individual observations.
The dashed black line has a slope of 0.4 and indicates that the faint end is sky dominated.
The distribution has been normalised along the magnitude axis, i.e. scaled so that each magnitude bin has the same number of sources in order to show features along the whole magnitude range. The greyscale is linear.}
\label{Fig:AccumGErr}
\end{figure}

The distribution of quoted error on the weighted mean\footnote{ The quoted error on the weighted mean includes a contribution
from the measured scatter and thus accounts for any underestimation of the errors on the individual transits. See the section on the ``Reference photometry update'' 
in \citet{PhotPrinciples}.} 
as a function of magnitude shown in \citet{PhotTopLevel} cannot be compared with expectations because each source has a different number of
observations. Figure~\ref{Fig:AccumGErr} shows the same analysis, but is restricted to those sources with between 90 and 110 CCD transits. 
The results for these sources 
can then be compared with predictions for
$N_{obs}=100$ using the formulation given in \citet{Jordi2010}. The lower line (green) gives the expected errors for a nominal mission and no calibration errors.
The zig-zag variation at G$<12$ shows the effect of gating which changes the effective exposure time of the observations. Adding a 3 mmag calibration error to this formulation shows the 
general level of calibration that has been achieved for \GDR1. 

Further features can be seen in this figure.

At the faint end, the main difference between the nominal and current mission is the increased stray light level which leads to poorer performance than expected. This cannot be calibrated out since
it is purely an increase in the noise level.

The jumps at approximately G=13 and G=16 are due to changes in the window class which affect the IPD algorithm and the number of pixels present in the image window. 
It should be noted that they do not occur exactly at these magnitudes since the plots are made with calibrated photometry, 
which is different to the   on-board  magnitude estimates that were used to determine the configurations (gate and window class) for each observation.

The increase in error seen at G=16 is linked to the change in the size of the window configuration. 
In the range 16 $<$ G $<$ 17, a limit is reached in the accuracy, probably caused by IPD issues. 

At G=13, the window class changes from 1D to 2D windows for the brighter transits and the IPD algorithm therefore changes \citep{IdtRef}. 
The greatest effect is that an AC LSF
component is needed in the fitting. At this early stage of the mission, the   best AC LSF to use is not very sophisticated and does not include colour or AC velocity dependencies.
Although the colour will remain the same for each observation for most sources, the AC velocity will not. This means that an additional noise is introduced into the flux determination.  
{ Moreover}, at this point the effect of flux loss affects observations fainter than G=13. When initialising the photometric system from raw observations, care must be taken to make sure that
discontinuities are not introduced into the system. This is described further in   Sect. 4 in \citet{PhotPrinciples}. If there are problems with this calibration,
then a larger scatter will be seen for sources around this magnitude.

At the bright end some of the increased scatter is caused by saturation.  Setting the gate configuration at the time of observation should remove most of the saturation
by changing the effective exposure time; however,  the accuracy of the on-board determination of the source magnitude, which determines the gate configuration, is poor (about 0.3 mag) at the bright 
end \citep{GaiaMission}. 
This means that
some observations are carried out with a gate configuration that does not eliminate saturation. While some masking of saturated pixels is carried out by the IPD, the calibration library used 
for this purpose is an early version from the commissioning period which only accounts for numerical saturation. Updates of this library will be in place for the next release.

The other
variations at the bright end are also caused by the different gate configurations being set. This changes the effective exposure time for each observation which alters the amount of smearing caused
by the AC velocity. The amount of additional noise seen will depend on the AC LSF selected. At this stage of the mission, no variation in the AC LSF is made as a function of AC velocity
\citep{IdtRef}.

From the accumulated data for each source, a P-value can be calculated from the $\chi^{2}$ of the weighted mean flux calculation. This is defined as the probability that the transits that have 
been used in forming the weighted mean flux for each source are normally distributed about the mean according to their quoted errors, i.e. there is no additional source of noise, for example source variability.

The P-values can be calculated using the  equation
\begin{equation}
P=Q\left({(n-1)\over2},{\chi^2\over2}\right)
,\end{equation}
where Q is an incomplete gamma function \citep{NumericalRecipes} and $n$ the number of CCD transits.

If the quoted errors are accurate and representative, this is a very good way of detecting variability. However, this is not the case with the current G-band data 
and the quoted errors from the IPD do not account for model inaccuracies, such as using  an LSF that is too simple in the IPD fit. 
This means that the CCD-level transits would be seen as having an underestimated quoted error. A consequence of this underestimation is that almost all sources have a G-band 
P-value of 0.0 and are seen as variable. Although no direct rescaling of the individual photometric errors is carried out for \GDR1, the calculation of the error on the weighted mean flux does take into account the
scatter of the data and thus this error is realistic.

\begin{figure} \resizebox{\hsize}{!}{\includegraphics{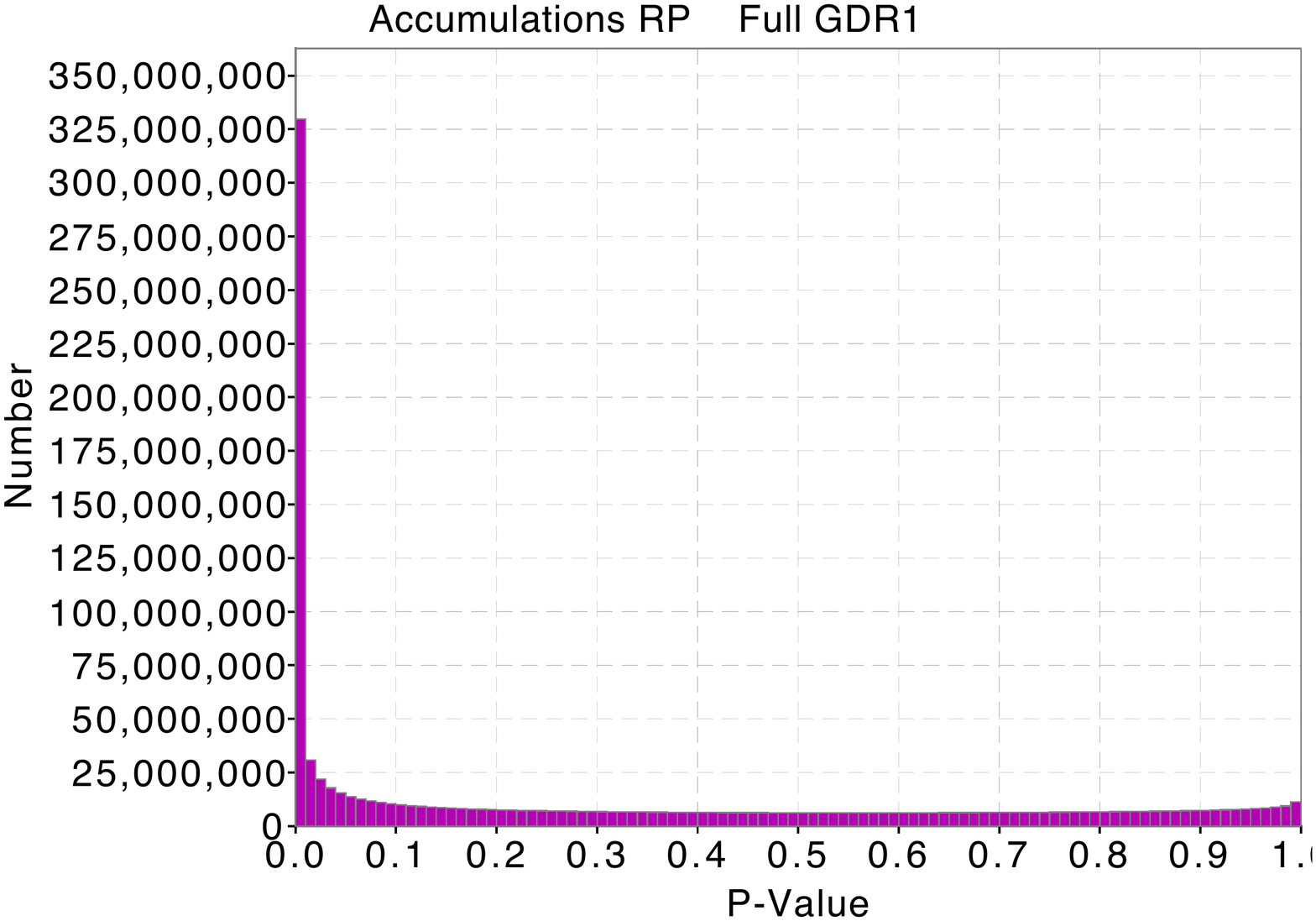}} \caption{ P-value distribution for \RP\ for the \GDR1 sources.}
\label{Fig:RpPValues}
\end{figure}

The situation is different for \BP\ and \RP\ since the flux determination is carried out using a simple integration rather than a model fit \citep{PhotPrinciples}. The errors here have contributions from
photon noise, background determination, and the geometric and differential dispersion calibrations.
As can be seen from Fig.~\ref{Fig:RpPValues}, the main feature in the P-value distribution for \RP\  is the peak at 0.0, which either indicates variability 
or that the calibration model is not well matched to the data for these sources. 
The significant flat distribution between 0.0 and 1.0 indicates that the quoted errors are realistic. No such flat distribution was seen in the equivalent G-band
analysis.

\section{Analysis of the residuals}

 A detailed analysis of the residuals allows us to validate the correctness of the calibration models by showing that
there are no systematic dependencies left from the calibration parameters after the application of the calibrations.
In this case residuals are computed as the difference between the calibrated epoch magnitude and the reference magnitude
for each source.

Each calibration unit is calibrated independently and therefore will naturally have residuals
centred on zero. We have analysed residual distributions for all CCDs in various
magnitude ranges, and indeed cannot see significant differences. 

\begin{figure}
\resizebox{\hsize}{!}{\includegraphics{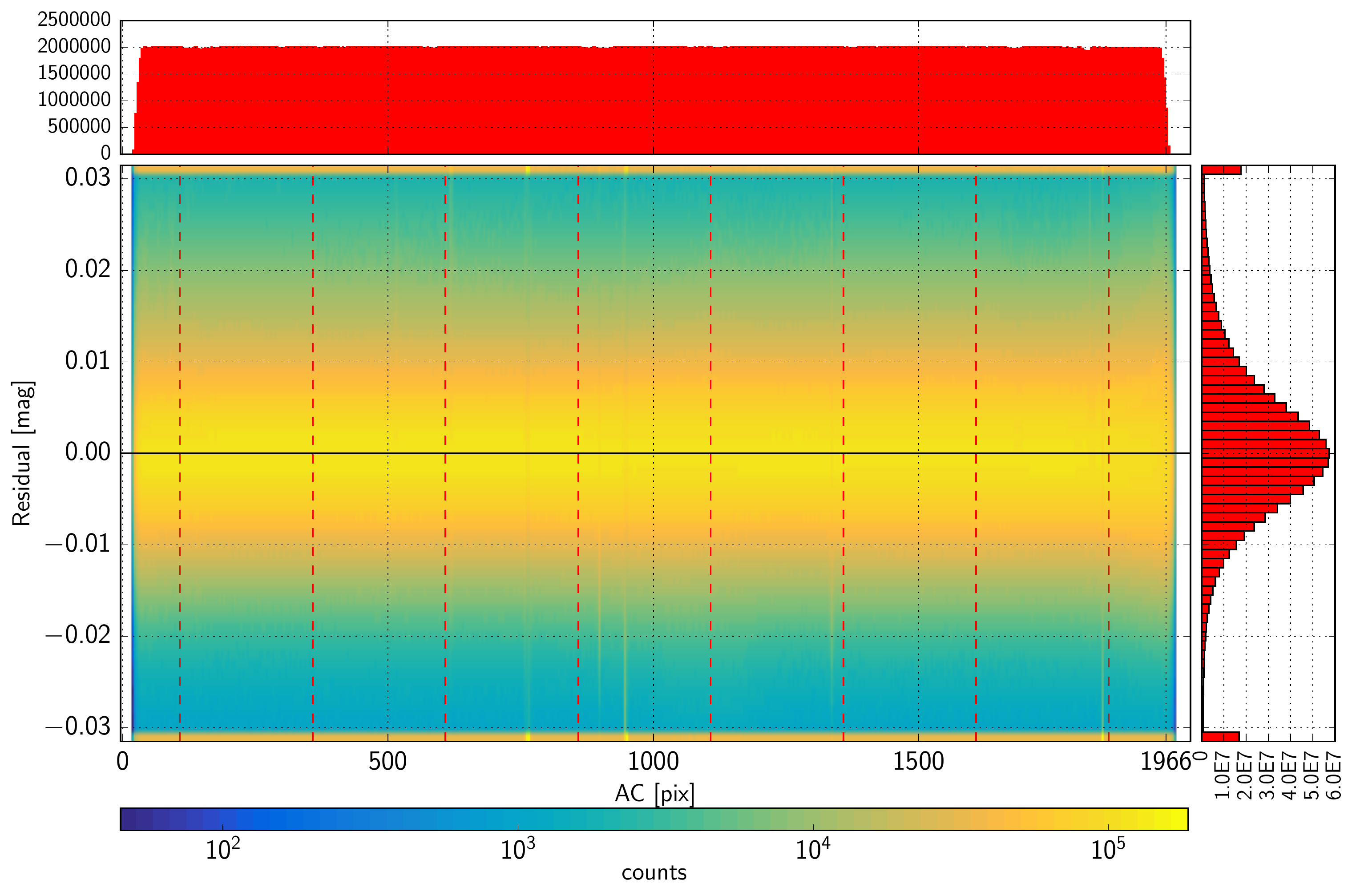}} \caption{Distribution of photometric residuals
against the AC coordinate for all data in AF1 CCDs and Window Class 1 (assigned to sources with magnitude 
$13<G<16$ as estimated on board).}
\label{Fig:Af1ResVsAc}
\end{figure}

In particular, residuals do not show any significant dependency on the calibration parameter AC coordinate. Figure
\ref{Fig:Af1ResVsAc} shows one such distribution (for the case of the AF1 CCDs and for the window class configuration
nominally assigned to sources with magnitude $13<G<16$). This is representative
of similar distributions in other locations on the focal plane.

\begin{figure}
\resizebox{\hsize}{!}{\includegraphics{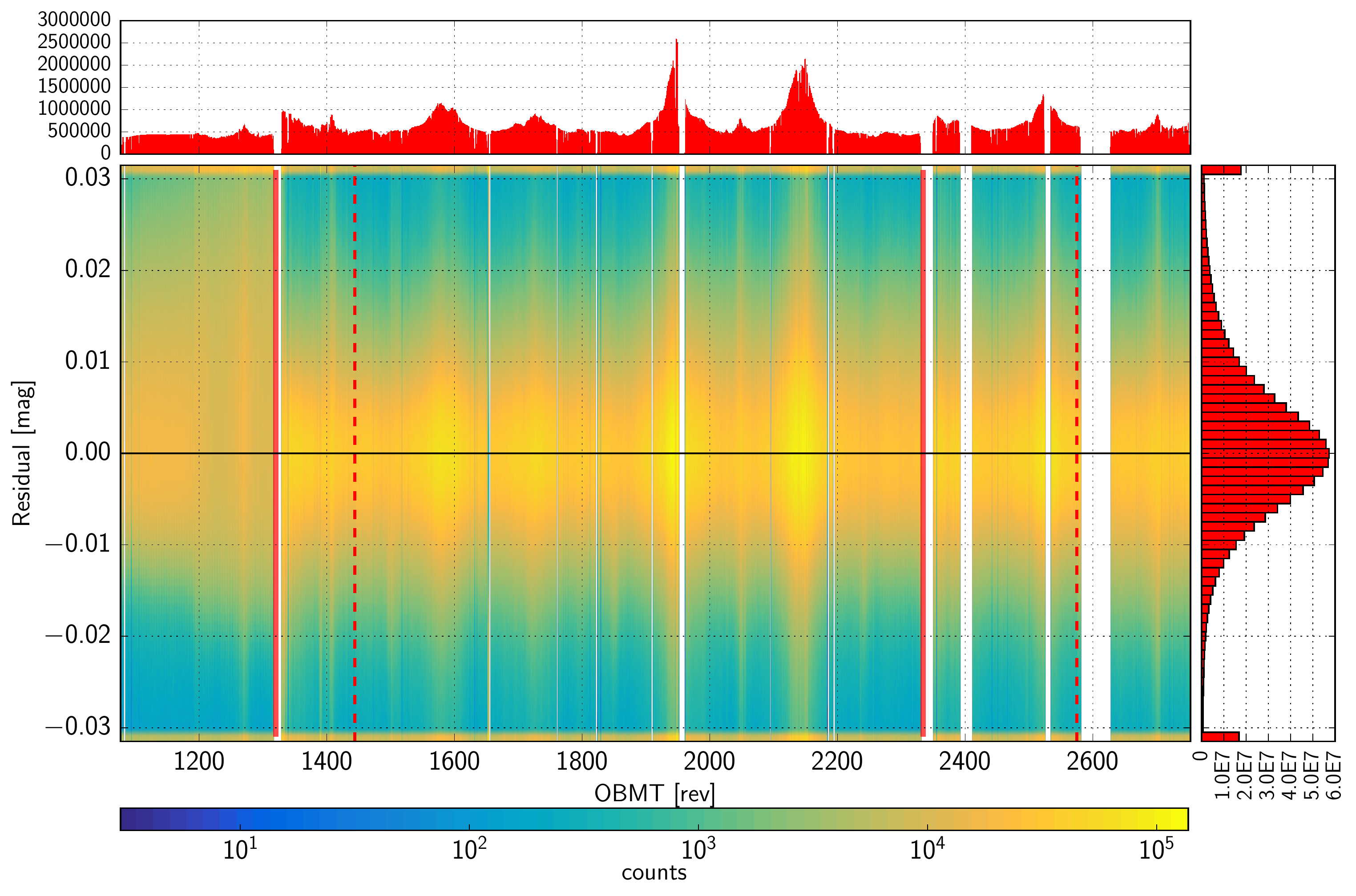}} \caption{Distribution of photometric residuals
in time for all data in AF1 CCDs and Window Class 1 (assigned to sources with magnitude in the
range $13<G<16$ as estimated on board). The time is given in OBMT revolutions (one revolution corresponds to  approximately  
6 hours). Vertical solid red lines mark the occurrences of decontamination activities, while dashed lines correspond to
refocus events.}
\label{Fig:Af1ResVsTime}
\end{figure}

In Fig. \ref{Fig:Af1ResVsTime} the distribution in time of the residuals for the same CCDs and magnitude 
range used in Fig. \ref{Fig:Af1ResVsAc} shows a non-Gaussian distribution of the residuals for the EPSL period where the data was heavily affected by contamination and poor LSF calibrations. From the first decontamination (marked by the first continuous 
vertical red line) onwards there is no sign of systematic problems in the residual distribution.

\begin{figure}
\resizebox{\hsize}{!}{\includegraphics{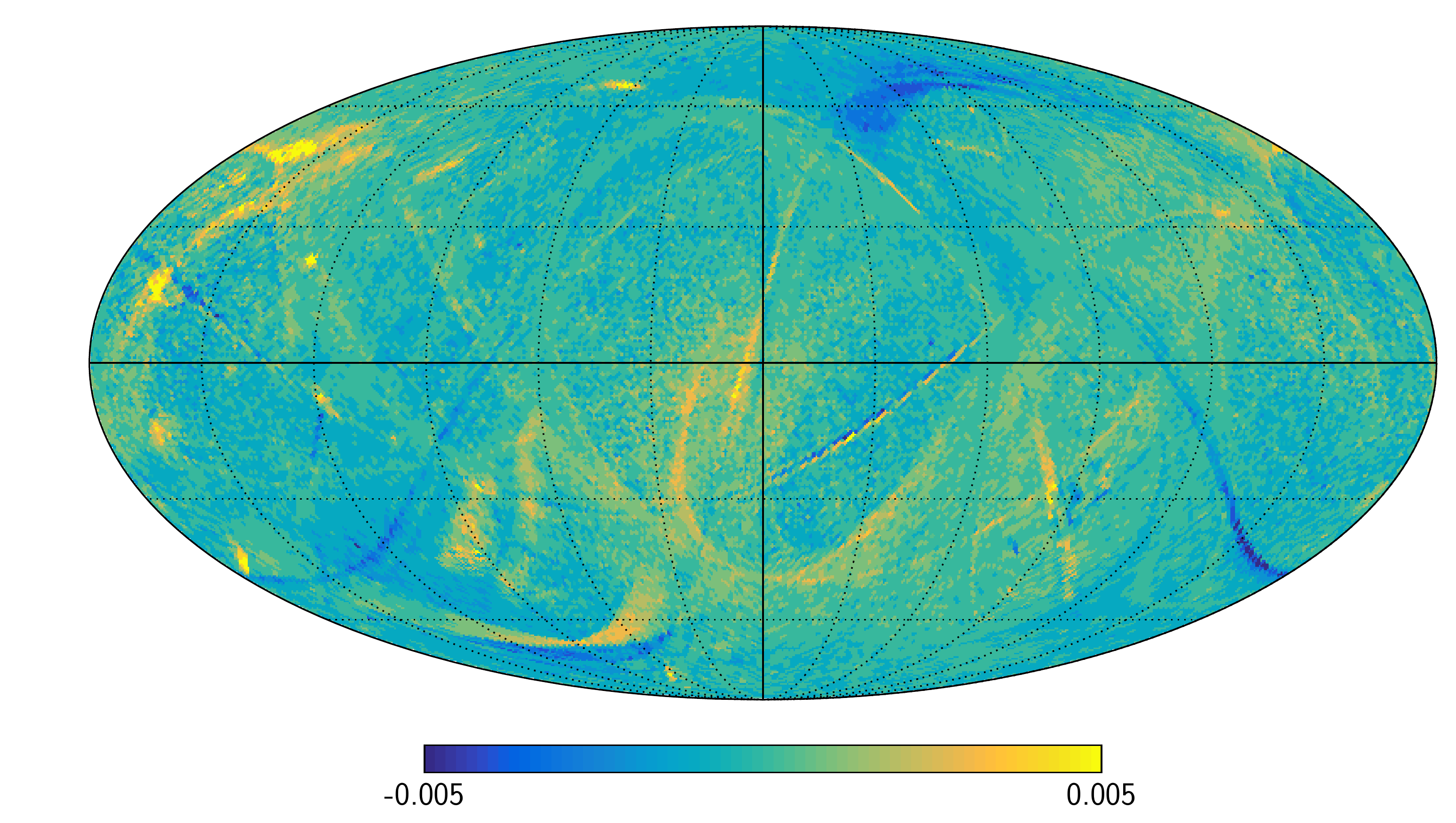}} \caption{Distribution of the median photometric 
residual in the sky for all data in AF1 CCDs and Window Class 1 (assigned to sources with magnitude in the
range $13<G<16$ as estimated on board). The map is shown using equatorial coordinates in Mollweide projection.}
\label{Fig:Af1ResSkyMap}
\end{figure}

A sky map of the median photometric residual (as shown in Fig. \ref{Fig:Af1ResSkyMap} for the same set of 
observations used in other residual plots in this section) indicates that there are some areas of the sky and
in particular some satellite scans that were not properly calibrated at the 0.01 mag level in the worst cases. 

\section{Analysis of data for mainly constant sources}\label{Sect:Constant}
One of the best ways of verifying the accuracy of the photometry is to carry out an analysis of constant sources. This makes the assumption that there is a population of
non-variable sources and that they can be selected such that they do not bias the results. Various studies using Hipparcos and Kepler data have shown that all stars are variable at some level,
but this is at a very low level \citep{KeplerVariability,EyerGrenon}. 

Given a selection of constant sources, by measuring their observed scatter the accuracy of the photometry can be assessed and compared to expectations. For constant sources, the scatter 
is caused by the two factors affecting the accuracy of the photometry, random noise and calibration error.
Because all sources are variable to some extent, the measured scatter will have a natural minimum value.
 \citet{KeplerVariability} showed that this was at the mmag level, but quite complicated in its dependency on
stellar type. 
Since this gives a minimum value in a similar way to that of an unknown calibration error, it is very difficult to distinguish between the two. Comparisons between the results in G, \BP,\ and
\RP\ can provide some information that can be helpful in distinguishing
 between the intrinsic variability of all sources and the calibration noise.

In the main part of the analysis of the photometry of these sources, a robust estimate of the scatter is made using the interquartile method mentioned earlier to estimate the standard deviation of the distribution. This is unaffected by photometric outliers which are likely to be present in this early reduction of the data.

As mentioned, care must be taken in selecting the constant sources. If the same scatter is used to exclude variables and to estimate the photometric accuracy, then the measured distribution will be 
biassed 
and narrower than it truly is. In order to avoid this bias,  it was decided that the analysis would be carried out on all sources since the majority of sources ($>90\%$) do not have large
amounts of variability \citep{EyerGrenon}.

This particular analysis is restricted to sources in the most observed sky regions (which have a mean of about 300 CCD transits).
In general, this restricts the sources to the ecliptic poles and the areas around ecliptic latitude $+45^{\circ}$ or $-45^{\circ}$.
These are areas that have been observed more often due to the scanning law of \Gaia.
{ A further random selection was carried out to reduce the number of faint sources such that there was a relatively flat distribution across the magnitude range.}

\begin{figure} \resizebox{\hsize}{!}{\includegraphics{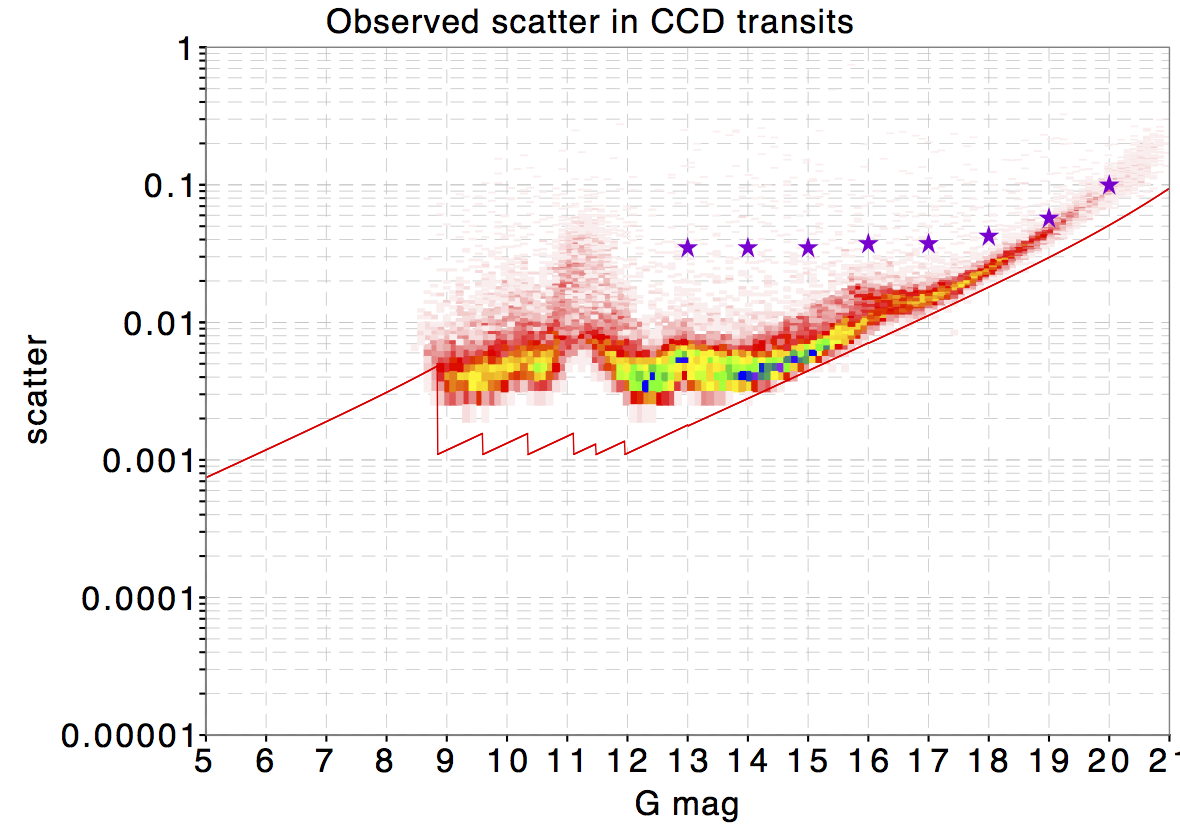}} \caption{Measured scatter in G for sources in the most observed sky regions as a function of magnitude. The line 
shows the predicted accuracies using the formulation given in \citet{Jordi2010}. The star symbols are the expected accuracies given on the \Gaia\ scientific performance web page. }
\label{Fig:ConstG}
\end{figure}

Figure \ref{Fig:ConstG} shows the results from this analysis for the G band as a function of magnitude. Also shown in this plot are the expected accuracies as derived from the equations
given in \citet{Jordi2010}. The discontinuities are due to different observing configurations such as gates and window class, which are controlled
as a function of magnitude as measured on board the satellite \citep{IdtRef}.
Between G=16 and 17, there seems to be a plateau in the scatter distribution pointing to
a possible accuracy limit.
For sources brighter than G=16, additional samples are transmitted by \Gaia, the accuracy improves and is closer to the expected values. 
Other features can be seen at the bright end and are consistent with the features seen in the results from the accumulations (see Fig.~\ref{Fig:AccumGErr}). The plot in Fig. 22  shows the scatter 
on CCD transits, while Fig.~\ref{Fig:AccumGErr} shows the error on the weighted mean for sources with about 100 CCD transits and thus accounts for the factor of 10 difference between them.

\begin{figure} \resizebox{\hsize}{!}{\includegraphics{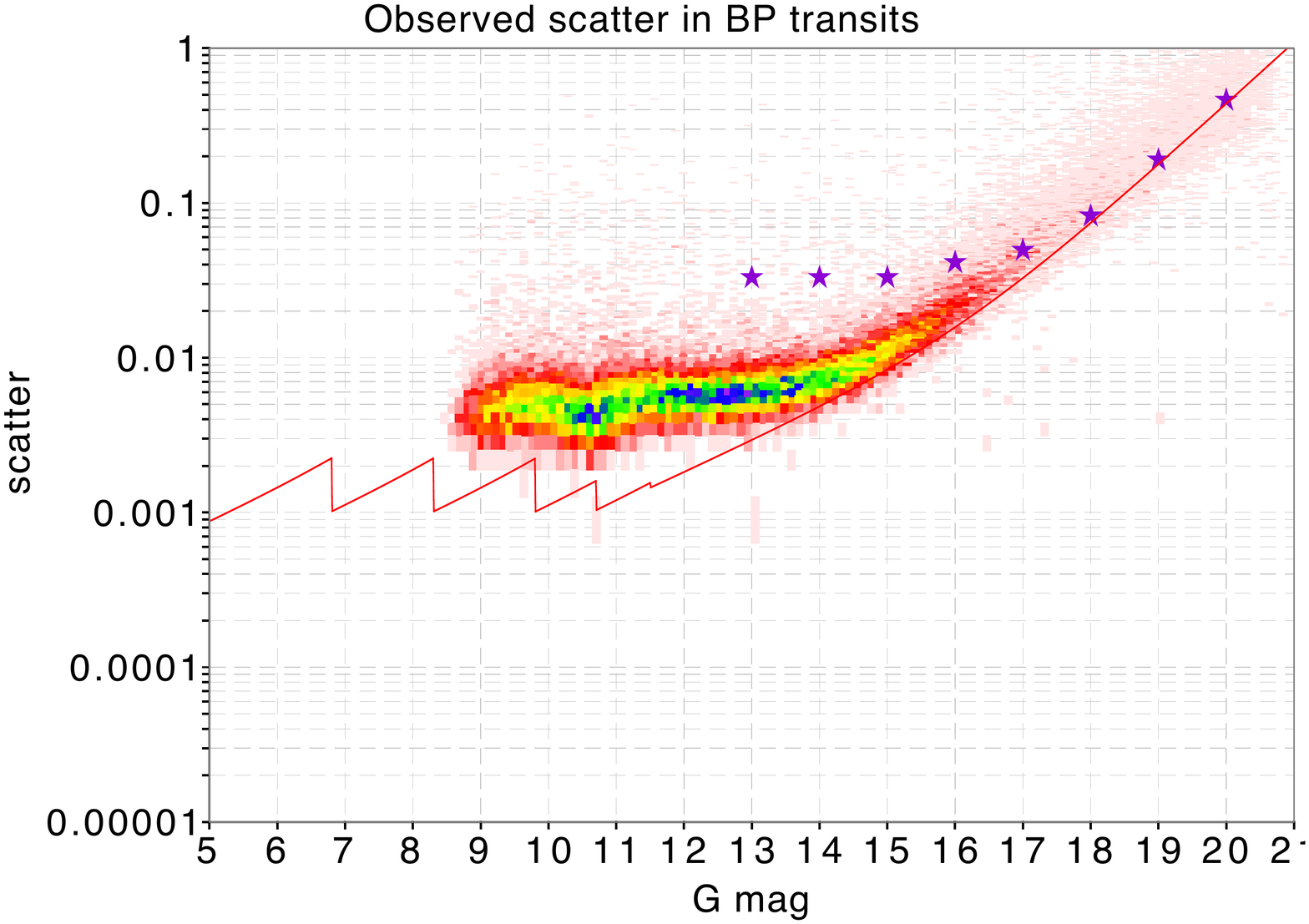}} \caption{As  Fig. \ref{Fig:ConstG}, but for \BP. }
\label{Fig:ConstBp}
\end{figure}

\begin{figure} \resizebox{\hsize}{!}{\includegraphics{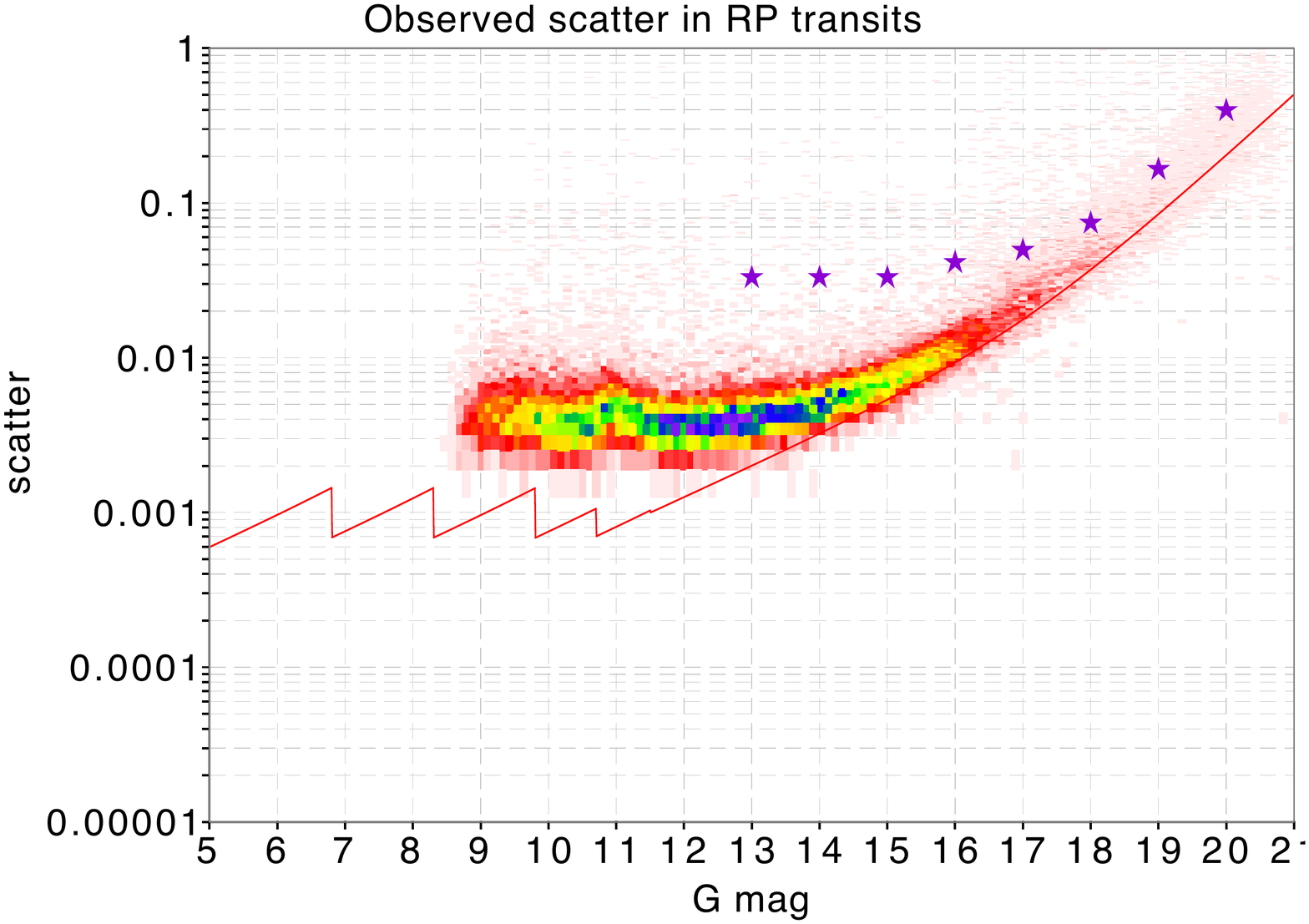}} \caption{As Fig. \ref{Fig:ConstG}, but for \RP. }
\label{Fig:ConstRp}
\end{figure}

 Toward faint magnitudes, the distribution of the data seems to follow  sky-dominated Poisson statistics at a level 
higher than expected. This is probably due to residual problems related to the stray light calibration for the G-band
measurements. Results at the faint end for BP and RP (see Figs. \ref{Fig:ConstBp} and \ref{Fig:ConstRp}) show a 
much better agreement with the expectations, thus confirming that the stray light calibration has successfully removed 
the effect of this additional background component on the integrated BP and RP photometry. It is worth reminding the reader that 
the G-band data enters the photometric calibration process in the form of image parameters, where the background 
calibration has already been computed and applied upstream \citep{IdtRef}, while for BP and RP the raw data is used.

Figure~\ref{Fig:ConstG} also shows the expected accuracies as given on the \Gaia\  scientific performance web 
page\footnote{\url{http://www.cosmos.esa.int/web/gaia/science-performance\#photometric\%20performance}}. These values reach a limiting accuracy at the bright end and are
due to assuming a calibration error of 30 mmag. From this plot, it can be seen that a better performance has already been achieved and the limiting accuracy is about 3 mmag.
This is still higher than the expected accuracy  due to photon noise  and it is foreseen that the performance will improve in future data releases as a better IPD is carried out 
and more complexity is added to the calibrations.

Figures \ref{Fig:ConstBp} and \ref{Fig:ConstRp} show the results for \BP\ and \RP. The limiting accuracies reached for these passbands are 3--4 mmag, similar to the G-band value. It should also  be noted that
these passbands are less affected at the bright end by saturation effects.

\begin{figure} \resizebox{\hsize}{!}{\includegraphics{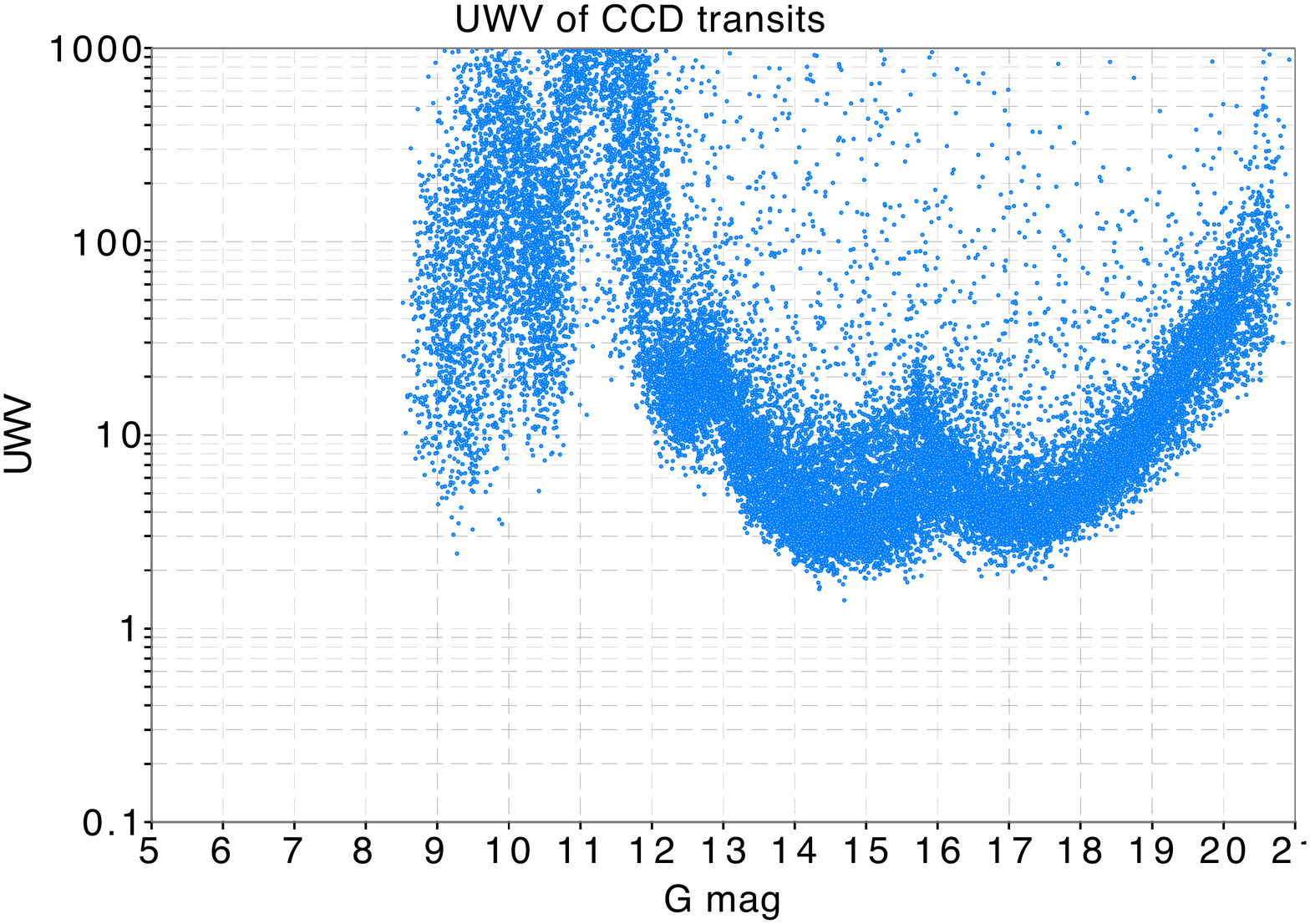}} \caption{Unit-weight variance as a function of magnitude for G.}
\label{Fig:GUwv}
\end{figure}

 An 
analysis that can be done on the same sample of sources to validate the errors estimated by the IPD \citep{IdtRef} is to investigate the unit-weight variance (Fig.~\ref{Fig:GUwv}). In this case, 
the variance or scatter is calculated with respect to the quoted error for each observation. 
The expected distribution should be centred around 1.0. This is not the case for this sample of sources  as  can be seen from Fig.~\ref{Fig:GUwv}. Large unit-weight variances can be due to
variability of the source, uncalibrated systematics, or  underestimated quoted errors. Results from previous analyses, in which comparisons were carried out with respect to expected
accuracies, show that there is not much additional scatter (due to variability or uncalibrated systematics).  Therefore,  we conclude that the quoted errors on the CCD-level transits are underestimated.
It is expected that this will improve with later IPD { results} in future processing cycles thanks to a better determination of the LSFs. However, the issue  may not be fully resolved
for the very brightest sources.

\section{Comparisons with external catalogues}\label{Sect:External}

\begin{figure*}
\centering
\includegraphics[width=\hsize*4/9]{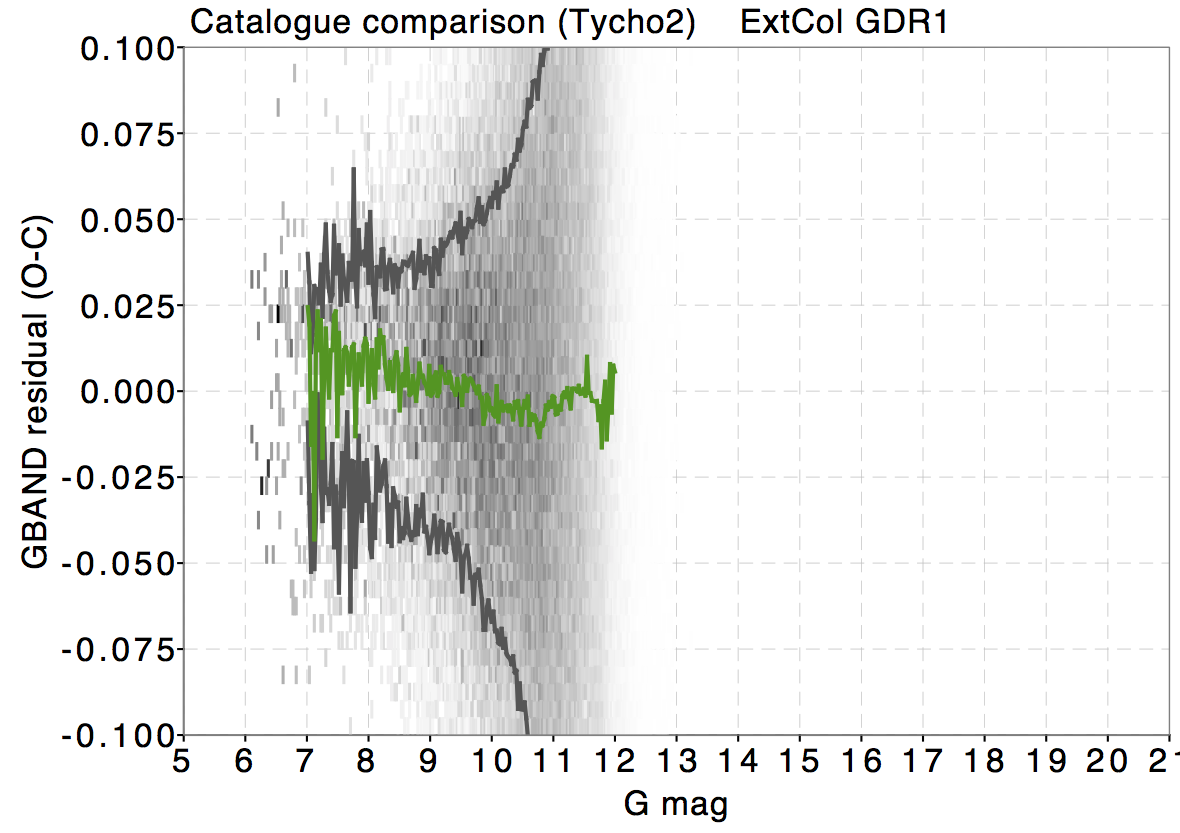} 
\includegraphics[width=\hsize*4/9]{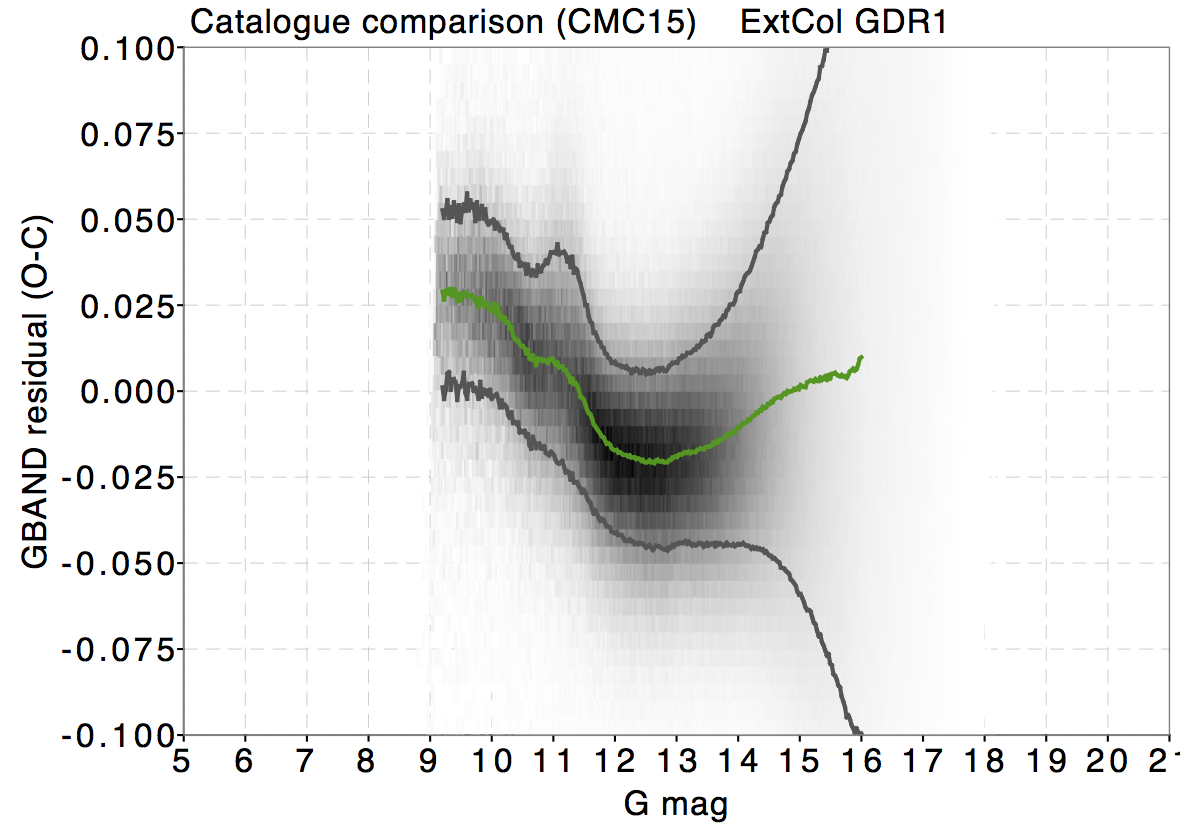} \\
\includegraphics[width=\hsize*4/9]{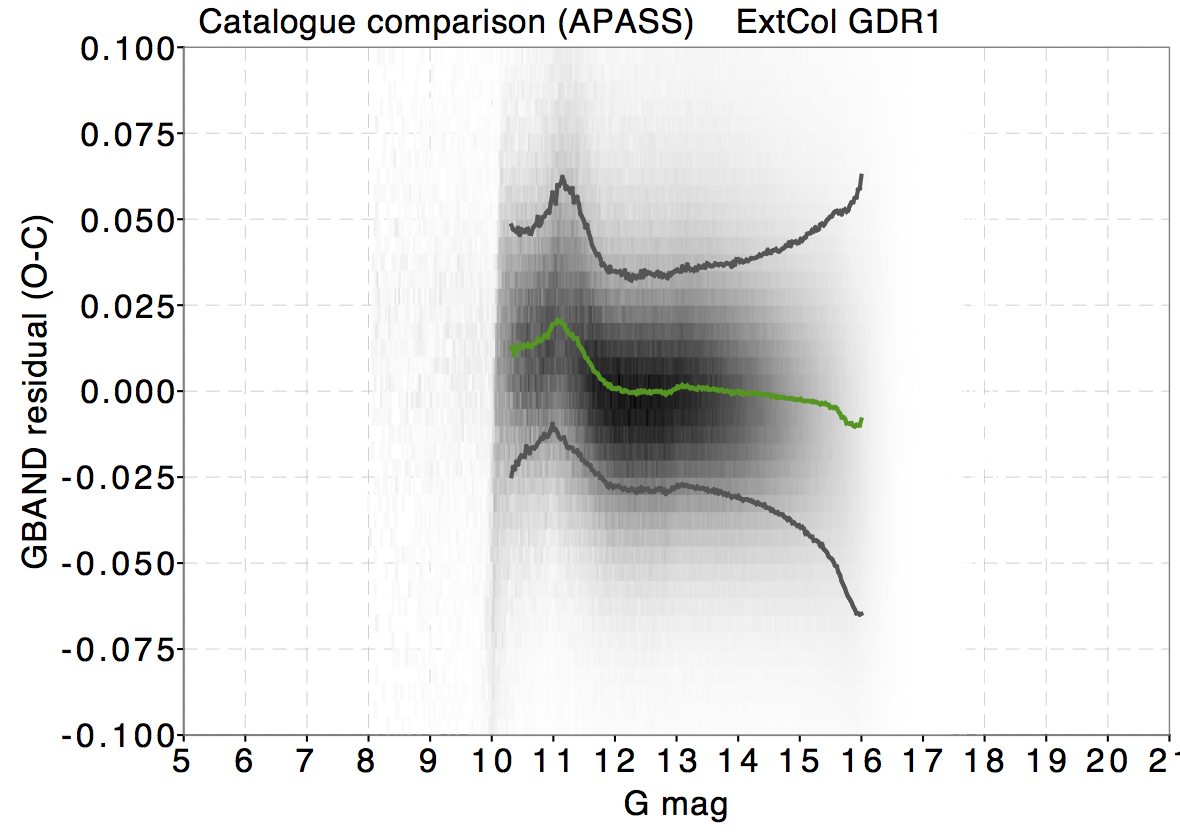} 
\includegraphics[width=\hsize*4/9]{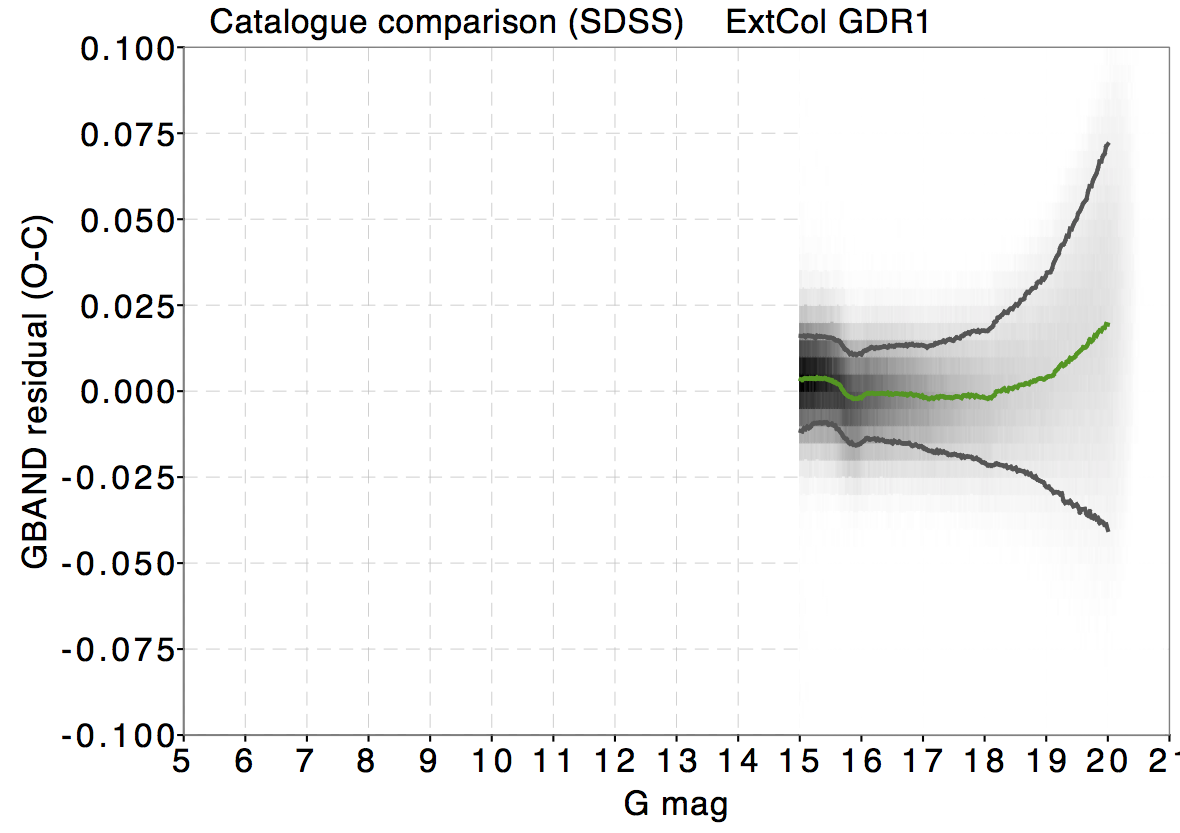} 
\caption{Comparisons with respect to the external photometric catalogues: Tycho 2, CMC15, APASS, SDSS.  The comparisons were carried out with respect to
the $r'$ passbands in all cases except for the Tycho comparison where the V$_{\rm T}$ passband was used.
The green and black lines show respectively the median and one sigma points of the residual distributions. The greyscale is linear.} 
\label{Fig:ExtComp}
\end{figure*}

Another way of validating results is to compare the data with other catalogues. The implied assumption with this { approach} is that the external catalogue is more reliable in some way than the data under test.
For \Gaia, this is usually not the case. The accuracy expected from the photometry will be much better than most external catalogues. Most of these catalogues will also not be all-sky and will contain
systematic errors of some sort. Another issue that complicates comparisons is that the angular resolution for \Gaia\  is much better than seen in any ground-based catalogue. This will cause many outliers
 when comparing areas of high source density such as the Galactic plane. For this reason, sources with $|b|<10^{\circ}$ are excluded from this analysis.

Comparisons have been done with respect to the following external photometric catalogues: Tycho 2 \citep{Tycho2}, CMC15 \citep{CMC15}, APASS \citep{APASS}, and SDSS DR12 \citep{SDSSDR12}.
The results of these comparisons are shown in Fig.~\ref{Fig:ExtComp}. 
In order to perform such comparisons, the passband of the external catalogue needs to be converted to the one used by Gaia. This is complicated by issues of absorption and luminosity
class. This  problem is avoided by restricting the colour range of the comparison  and by positioning this narrow range such that the difference between G and the passband being compared is constant, 
thus making a direct comparison possible.
The range also has to be selected such that there is a reasonable population of sources present to analyse.
In this comparison, the range chosen was 1.0$<$\BP$-$\RP$<$1.2. A single zeropoint offset is then determined for each catalogue to facilitate comparison.

 The analysis of the distribution widths of these comparisons to validate the quoted errors is unreliable 
and difficult to interpret; the quoted errors of the external catalogues often refer to different magnitude ranges
than the ones where the comparison distributions are narrowest.
The quoted errors on the external catalogue are reasonably close to the minimum standard deviation in the comparisons,
but no firm conclusions can be drawn from this.

The four catalogues chosen cover different magnitude ranges. The most useful catalogues to
use are CMC15 and APASS since they approximately cover the same magnitude range and are of similar accuracies.  When anomalies are seen in the comparisons, it is difficult to
assess  whether the problem  lies in the \Gaia\  photometry or  in the external catalogue. 
 If an anomaly is  seen in more than one external catalogue comparison, it is likely that the issue is with the \Gaia\  photometric results. This is especially
the case if a  plausible cause can be found. During the early reductions of the data, a jump was found at G=13 in both these comparisons which corresponds to a window class change in \Gaia. 
This anomaly was dealt with by the Gate/Window Class link calibrations. See \citet{PhotPrinciples} for more details. The plots in Fig.~\ref{Fig:ExtComp} show that this issue has been resolved.

These two comparisons also show one of the problems linked with this type of analysis. At the bright end of the CMC15 comparison there seems to be a magnitude term. Since this is
not seen in the APASS results, this difference is likely to be in the CMC15 photometry. A reason for this deviation could be  the asymmetrical images that occur away from the equator and
that might cause problems for the isophotal corrections carried out in this catalogue \citep{CMC14}. However, the bump of order 0.01 mag at G=11 is likely to be in the \Gaia\  data 
and might be caused by saturated images (also see Sect.~\ref{Sect:Accumulation}).

Although the accuracy of the Tycho 2 photometry is much worse  than that of \Gaia, a comparison with that catalogue is useful as it covers the brighter sources and shows rather good linearity.
The SDSS comparison checks the fainter end of the magnitude range and shows
that there are no large-scale anomalies. 
The slightly positive differences at the faint end can be  real trends or can be incompleteness of the \Gaia\  data due to the magnitude detection threshold. 
It should be noted that the SDSS  comparison is limited to levels  fainter than $G=15$ owing to saturation effects in the SDSS photometry \citep{SdssSaturation}.

\begin{figure*}
\centering
\includegraphics[width=\hsize*5/9]{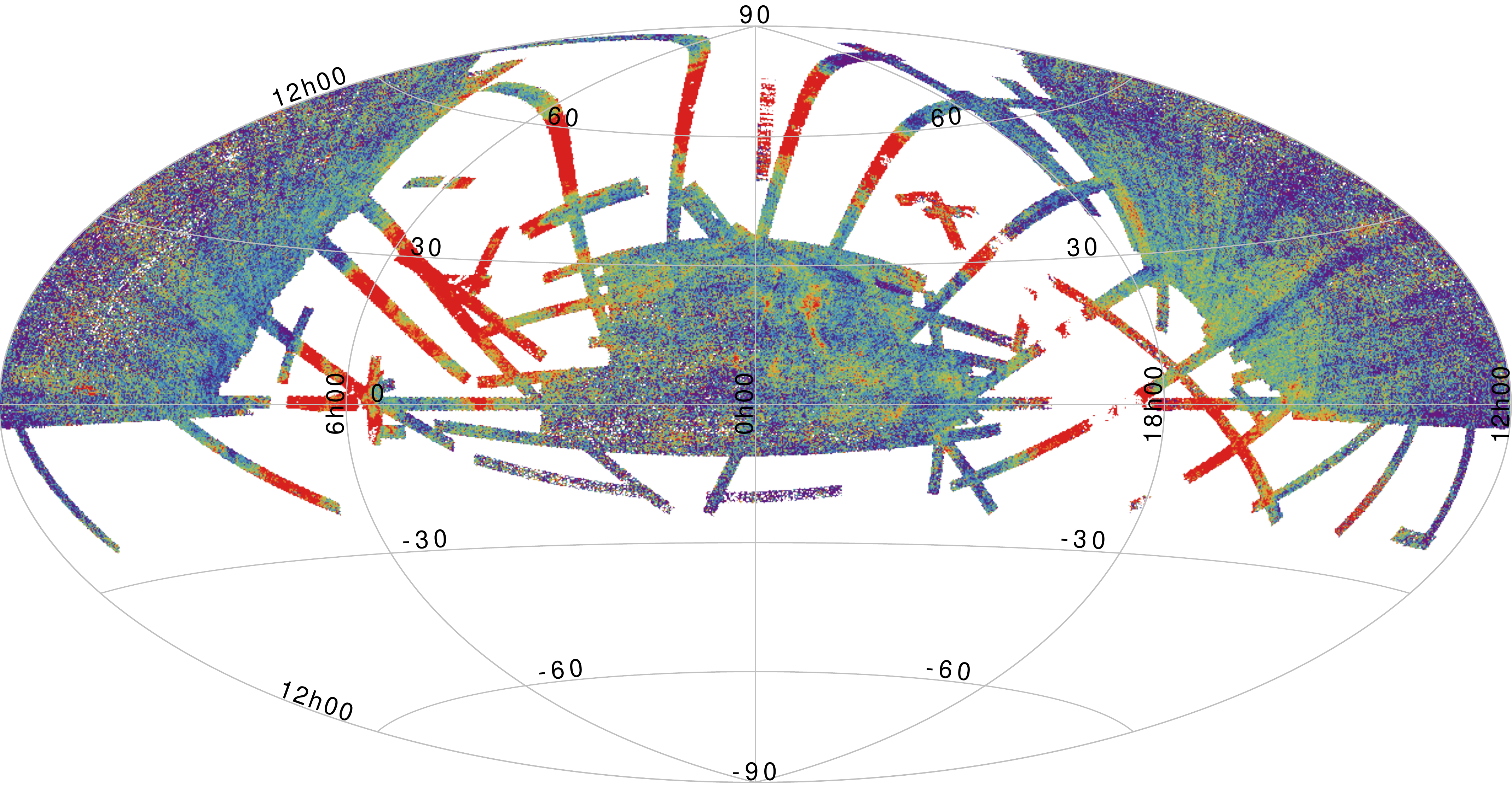} 
\includegraphics[width=\hsize*3/9]{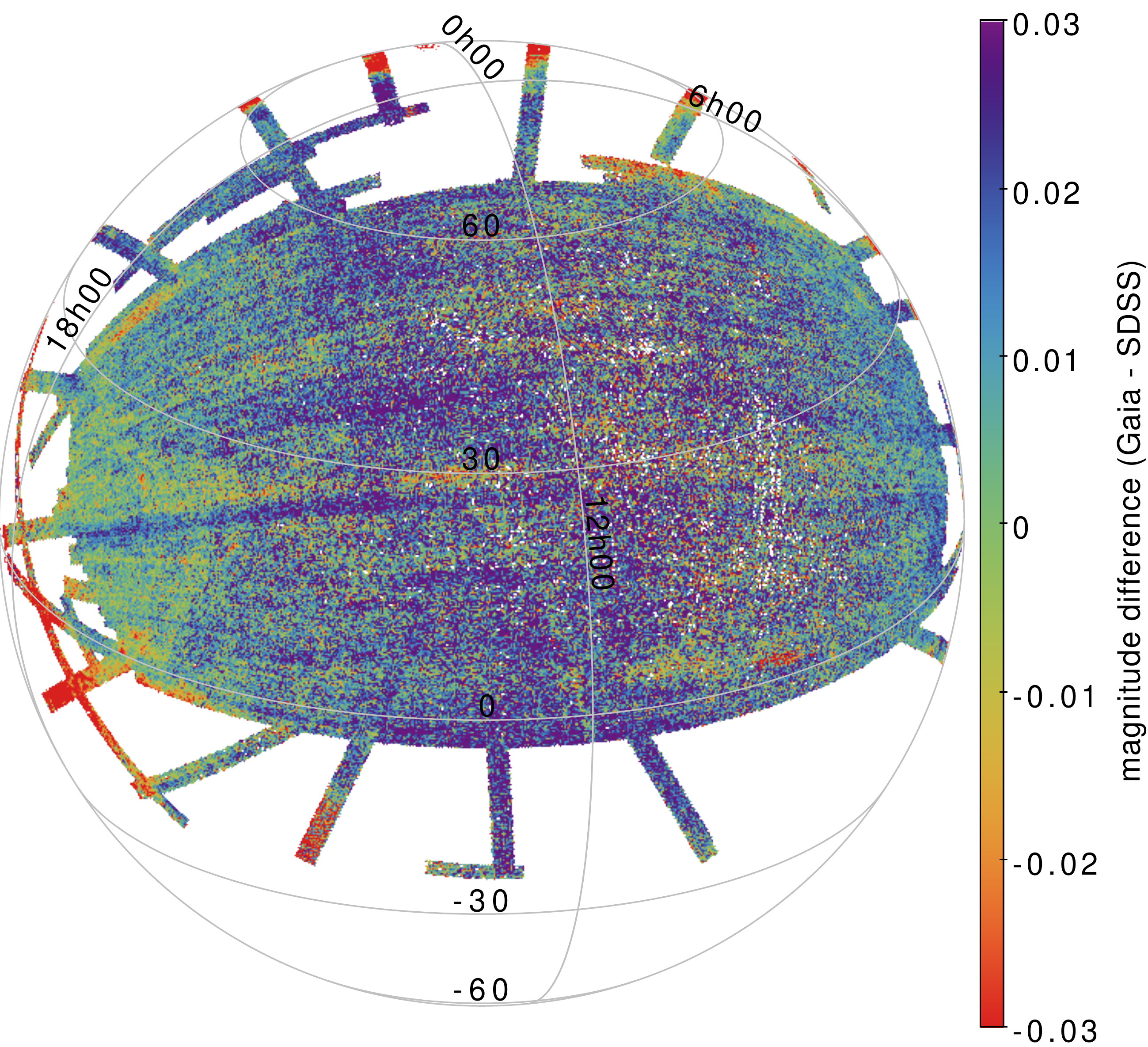} 
\caption{ Sky distribution of the median zeropoint between the \Gaia\  and SDSS photometric catalogues. The comparison
is limited to sources fainter than $G=15$ and to sources in the colour range $0.8<g-r<1.1$. The empirical photometric
transformations from \citet{PhotTopLevel} have been used to estimate $G$ magnitudes from $g$ and $r$. The entire sky in 
equatorial coordinates is shown on the left, while the plot on the right shows a 3D view centred on the large area outside 
the Galactic plane covered by SDSS.}
\label{Fig:SdssSky}
\end{figure*}

{ A plot of the sky distribution of the magnitude zeropoint between the \Gaia\  and SDSS catalogues is shown in 
Fig. \ref{Fig:SdssSky}. In order to have a good coverage of the regions of the sky observed by SDSS, the
selection in colour has been relaxed to $0.8<g-r<1.1$ and the empirical photometric transformations from 
\citet{PhotTopLevel} have been applied to minimise the colour effects. Figure \ref{Fig:SdssSky} shows  the entire sky in equatorial coordinates
in Hammer-Aitoff projection (left) and  a view of the large area outside the Galactic plane covered 
by SDSS (right). The Galactic plane stands out with larger differences between the \Gaia\  and SDSS magnitudes. Outside the 
Galactic plane the most prominent feature is the SDSS scanning pattern thus showing that the SDSS photometry
dominates the error budget.}

\section{Conclusions}\label{Sect:Conclusions}

This paper has described the internal validation investigations on the photometry carried out for the first \Gaia\  data release. Although only G-band photometry is present in
\GDR1, some validation of the \BP\ and \RP\ photometry is shown since it is used in some of the G-band calibrations.
In general, the photometric calibrations have been carried out to the 3--4 mmag level, but there are systematics at the 10 mmag level at G=11.

\begin{acknowledgements}

This work has been supported by the UK Space Agency, the UK Science and Technology Facilities Council.

The research leading to these results has received funding from the 
European Community's Seventh Framework Programme (FP7-SPACE-2013-1) 
under grant agreement no. 606740.

This work was supported in part by the MINECO (Spanish Ministry of Economy) - FEDER through grant ESP2013-48318-C2-1-R and MDM-2014-0369 of ICCUB (Unidad de Excelencia `Mar\'ia de Maeztu').

We also thank the Agenzia Spaziale Italiana (ASI) through grants ARS/96/77, ARS/98/92, ARS/99/81, I/R/32/00, I/R/117/01, COFIS-OF06-01, ASI I/016/07/0, ASI I/037/08/0, ASI I/058/10/0, ASI 2014-025-R.0, ASI 2014-025-R.1.2015,
and the Istituto Nazionale di AstroFisica (INAF).

The work was supported by the Netherlands Research School for Astronomy (NOVA) and the Netherlands Organisation for Scientific Research (NWO) through
grant NWO-M-614.061.414.

This research has made use of the APASS database, located at the AAVSO web site. Funding for APASS has been provided by the Robert Martin Ayers Sciences Fund.\\

\end{acknowledgements}

\bibliographystyle{aa} 
\bibliography{refs} 

\begin{appendix}
\section{Nomenclature}\label{acronyms}
Below, we give a list of acronyms and useful concepts used in this paper. \hfill\\
\begin{table*}
\begin{tabular}{lp{0.35\textwidth}}\hline\hline 
\textbf{Acronym} & \textbf{Description} \\\hline
%
AC&ACross scan: direction  on the focal plane perpendicular to the scan direction \\\hline
AF&Astrometric Field:  the 62 astrometric CCDs on the focal plane\\\hline
APASS&AAVSO Photometric All-Sky Survey \\\hline
BP&Blue Photometer:  the system containing the blue dispersion prism. Also refers to the associated CCDs.\\\hline
CCD&Charge-coupled Device \\\hline
CCD transit&Transit of a source across a single CCD \\\hline
CMC15&Carlsberg Meridian Catalogue, Number 15 \\\hline
EPSL&Ecliptic Pole Scanning Law:  the scanning law of the satellite, where it is pointing as a function of time, during the first 
month of observations. See \citet{GaiaMission} for more details on the various scanning laws of Gaia. \\\hline
ESA&European Space Agency \\\hline
FoV&Field of View:  one of the two pointing directions of the satellite telescopes. See \citep{GaiaMission} for more information regarding the
structure of Gaia. \\\hline
FoV transit&Field-of-view transit, the complete transit of a source across the focal plane\\\hline
IPD&Image Parameter Determination:  The task that generates the fluxes that are calibrated as described in \citet{PhotPrinciples}. This task
also generates the raw astrometry. See \citet{IdtRef} for more details. \\\hline
LS&Large-scale: { usually referred to in the context of the relevant photometric calibration} \\\hline
LSF&Line Spread Function \\\hline
OBMT&On-board Mission Timeline:  the timescale usually used when referring to time in the Gaia context. This scale is
defined in \citet{GaiaMission} \\\hline
RP&Red Photometer:  the system containing the red dispersion prism. Also refers to the associated CCDs.\\\hline
SDSS&Sloan Digital Sky Survey \\\hline
SS&Small-scale:  usually referred to in the context of the relevant photometric calibration \\\hline
SSC&Spectrum Shape Coefficient:  equivalent of a medium-band colour. These are defined in \citet{PhotPrinciples}. \\\hline
TDI&Time-delayed Integration (CCD) \\\hline
\end{tabular} 
\end{table*}
\end{appendix}


\end{document}